\documentclass[twocolumn]{aastex62}

\usepackage{enumitem}

\shorttitle{Statistical Analyses of Solar Active Region Features}
\shortauthors{Chen et al.}
\begin{document}

\title{Statistical Analyses of Solar Active Region in SDO/HMI Magnetograms detected by Unsupervised Machine Learning Method DSARD}

\correspondingauthor{Q. Hao}
\email{haoqi@nju.edu.cn}

\author{R. Chen}\footnote{R. Chen and W. Lu are co-first authors.}
\affiliation{School of Astronomy and Space Science, Nanjing
	University, Nanjing 210023, China}
\affiliation{School of Mathematics, Nanjing
	University, Nanjing 210023, China}

\author{W. Lu}
\affiliation{School of Astronomy and Space Science, Nanjing
	University, Nanjing 210023, China}
%\affiliation{W. Lu and R. Chen are co-first authors.}

\author[0000-0002-9264-6698]{Q. Hao}
\affiliation{School of Astronomy and Space Science, Nanjing
	University, Nanjing 210023, China}
\affiliation{Key Laboratory of Modern Astronomy and Astrophysics
	(Nanjing University), Ministry of Education, Nanjing 210023, China}

\author{Y. Meng}
\affiliation{School of Astronomy and Space Science, Nanjing
	University, Nanjing 210023, China}

\author[0000-0002-7289-642X]{P. F. Chen}
\affiliation{School of Astronomy and Space Science, Nanjing
	University, Nanjing 210023, China}
\affiliation{Key Laboratory of Modern Astronomy and Astrophysics
	(Nanjing University), Ministry of Education, Nanjing 210023, China}

\author{C. Shi}
\affiliation{School of Astronomy and Space Science, Nanjing
	University, Nanjing 210023, China}



\begin{abstract}

 Solar active regions (ARs) are the places hosting the majority of solar eruptions. Studying the evolution and morphological features of ARs is not only of great significance to the understanding of the physical mechanisms of solar eruptions, but also beneficial for the hazardous space weather forecast. An automated DBSCAN-based Solar Active Regions Detection (DSARD) method for solar ARs observed in magnetograms is developed in this work, which is based on an unsupervised machine learning algorithm called Density-Based Spatial Clustering of Applications with Noise (DBSCAN). The method is then employed to identify ARs on the magnetograms observed by the Helioseismic and Magnetic Imager (HMI) onboard Solar Dynamics Observatory (SDO) during solar cycle 24 and the rising phase of solar cycle 25. The distributions of the number, area, magnetic flux, and the tilt angle of bipolar of ARs in latitudes and time intervals during solar cycle 24, as well as the butterfly diagram and drift velocities are obtained. Most of these statistical results based on our method are in agreement with previous studies, which also guarantees the validity of the method. In particular, the dipole tilt angles in ARs in solar cycle 24 and the rising phase of solar cycle 25 are analyzed which reveal that 13\% and 16\% of ARs, respectively, violate Hale's law. 

\end{abstract}



\section{Introduction} \label{sec:intro}

Active regions (ARs) are the patchy volumes in the solar atmosphere with strong magnetic fields, where a variety of activities occur, manifesting in complex spatial and temporal behaviors extending from the photosphere to the corona \citep{Driel2015}. Sunspots, being the initial features discernible as tracers of ARs on the Sun, appear dark areas on the surface of the Sun in continuum, occasionally perceptible to the naked eyes \citep{Solanki2003}. They mark the locations where magnetic field is sufficiently strong to suppress the convective flows that transport heat from the interior to the solar surface. Observations of sunspots in ARs have revealed several characteristics that are crucial to understanding the physical mechanisms of the solar cycles. One such example is the equatorward migration of sunspots known as the ``Butterfly Diagrams" \citep{Maunder1904}, which further led to the discovery of various magnetic properties of sunspots, including the famous Joy's law and Hale's polarity law \citep{Maunder1904}. The most significant aspect of ARs and sunspots are their capacity to generate strong solar flares and, sometimes, coronal mass ejections (CMEs), which have the potential to significantly impact the space weather. Consequently, the study of the evolution of ARs and their relationship with solar eruptions is not only useful in predicting eruptive events \citep{Schrijver2007, Chen2011, Bobra2015, Barnes2016, Sun2022, Zhang2022}, but also in understanding the long-term evolution of solar activities \citep{Hathaway2010, Ravindra2022}.

The National Oceanic and Atmospheric Administration (NOAA) Space Weather Prediction Center (SWPC) has developed a widely referenced catalog of active regions (ARs) based on white-light images. NOAA assigns a unique number to an AR if it contains at least one sunspot, with the numbering determined either manually or through computer-aided inspection assisted by human operators. This catalog offers detailed information, including the location of sunspot groups, their areas, and magnetic morphology.

However, significant advancements in observational techniques have led to a substantial increase in both the quantity and quality of multiwavelength observational data. Traditional manual detection or labeling of ARs on images has become increasingly inadequate in meeting the demands for rapid and accurate processing of large datasets, particularly for long-term statistical studies and solar dynamo modeling. Recognizing the limitations of human inspection, the development of automated detection methods has gained momentum. These methods aim to provide objective and consistent extraction of ARs, addressing the challenges posed by manual processes and enabling more efficient analysis of solar data.

An early class of automated detection methods for ARs was primarily developed using image processing techniques \citep{Zharkov2005, McAteer2005, McAteer2005b, Benkhalil2006, Colak2008, Zhang2010, Higgins2011, Cui2021}. \citet{Verbeeck2013} conducted a comprehensive comparison of four typical automated detection methods for ARs or sunspots observed in images with different wavelengths. These methods typically involve threshold segmentation on magnetograms or white-light images, followed by region-growing and morphological operations to derive the areas of ARs. For instance, \citet{Zharkov2005} employed an edge detection method after standardizing images to identify sunspot candidates. This was followed by local thresholding for image segmentation, and subsequent watershed and morphological operations to smooth features and fill gaps. The temporal variations in the detected sunspot areas showed a high correlation with NOAA's records, validating the effectiveness of the proposed method. This sunspot detection approach was applied to white-light images captured by the Michelson Doppler Imager (MDI; \citealt{Scherrer1995}) aboard the Solar and Heliospheric Observatory (SOHO) spacecraft from 1996 to 2004. The results revealed that the number of sunspots increases near-exponentially as the area decreases over the observed period. Additionally, the sunspot areas exhibited a periodicity of approximately 7–8 years in terms of north-south asymmetry.

\citet{Zhang2010} developed an automated AR detection method incorporating three image processing techniques: intensity thresholding, region-growing, and morphological operations. This method was tested and applied to the Carrington synoptic magnetograms constructed from SOHO/MDI images spanning 1996 to 2008. The method identified 1730 ARs and yielded significant statistical insights, including morphological features, magnetic flux, and drift velocities during solar cycle 23. The true positive and false positive rates relative to the NOAA AR catalog were found to depend on the detection threshold. For example, the frequency distribution of ARs in terms of area size and magnetic flux follows a log-normal function, but shifts to a power-law distribution when the detection thresholds are lowered. Despite computational biases, the authors suggested that the distribution of solar magnetic features, from large to small, might consist of two components: a log-normal distribution for large ARs and a power-law distribution for smaller ARs and other minor features. However, these methods rely on empirically assigned thresholds and struggle to effectively separate ARs in complex magnetic field regions on high-resolution images, leading to incomplete identification. Their robustness depends on thresholds determined through extensive trial and error, making them susceptible to instability due to detector sensitivity degradation. 

The advancement of machine learning methods in computer vision technology has significantly expanded their application in the detection of solar ARs. One widely used approach is based on clustering, a form of unsupervised machine learning. \citet{Barra2008} introduced a fuzzy clustering technique to automatically segment EUV images into regions of coronal holes, quiet Sun, and active regions. Noticing some artifacts of the results, they later improved the method and applied it to the entire set of EIT solar images over solar cycle 23 \citep{Barra2009},
 yielding results consistent with earlier studies. \cite{Caballero2013} applied clustering methods to group segmented regions into active regions for EUV observations via the region growing technique. A comparative study of three clustering techniques revealed that hierarchical clustering delivered the best performance.

\cite{Jiang2022} introduced a novel approach by first applying the scale-invariant feature transform (SIFT) to extract different regions. These regions were then clustered based on their density peaks, enabling the automatic detection of active regions in magnetograms. A key advantage of this method over the region-growing approach is that it eliminates the need for seed point selection, thereby avoiding the issue of multiple neighboring active regions being incorrectly identified as a single AR.

It is worth noting that \cite{Turmon2002} pioneered a Bayesian image segmentation technique, leveraging statistical models trained on scientist-provided labels to characterize the quiet Sun, faculae, and sunspots. This technique was further extended and refined in a subsequent study \citep{Turmon2010}. The method was applied to identify active region patches observed by the Helioseismic and Magnetic Imager (HMI; \citealt{Schou2012}) onboard the Solar Dynamics Observatory (SDO; \citealt{Lemen2012}). This led to the creation of the widely used Spaceweather HMI Active Region Patch (SHARP; \citealt{Bobra2014}), which provides detailed maps containing information on magnetic fields, Doppler velocity, and continuum intensity.

Convolutional neural networks (CNNs) represent another widely utilized machine learning method, standing at the forefront of feature segmentation in the machine learning domain. Supervised learning algorithms typically train models by adjusting parameters based on labeled data until the desired performance is achieved. \citet{Quan2021} employed two object detection models, Fast R-CNN and YOLO-V3, to detect the bounding boxes of ARs in magnetograms observed by SDO/HMI. The experimental results demonstrated that both methods exhibit high detection accuracy and rapid processing speeds. While bounding boxes around ARs may be effective for flare prediction, focusing on the detailed characteristics of ARs reveals that bounding boxes alone are insufficient. To accurately extract specific AR regions within these boxes, intensity threshold processing or pixel-scale classification, as proposed by \citet{Turmon2010}, is necessary. Semantic segmentation models, such as U-Net \citep{Ronneberger2015}, have proven effective in accurately identifying regions of interest for automatically detecting filaments/prominences and ARs with high precision \citep{Zhu2019, Liu2020, Liu2021, Zhang2024}. However, unlike other solar activities, ARs in magnetograms often appear as diffuse, fragmented areas of dots and patches. This indicates that while semantic segmentation models can delineate fragmented patches of ARs, an additional step is required to determine which fragments belong to a single AR. One potential solution is to employ an instance segmentation model, as previously implemented for filaments by \citet{Guo2022}. However, the number of AR fragments significantly exceeds that of filaments, complicating the data labeling process.

Despite the higher accuracy and generalization ability of these supervised learning models, they require large amounts of objective and consistent labeled data of ARs, which is difficult to free from the subjectivity of human inspection. In contrast, unsupervised learning algorithms do not require large amounts of data for feature extraction but instead on an appropriate algorithm according to the characteristics of the data itself. In this paper, we employ Density-Based Spatial Clustering of Applications with Noise (DBSCAN) algorithm, a classical density-based spatial clustering unsupervised learning algorithm \citep{Schubert2017},  to develop an automated detection method for solar active regions observed in magnetograms. The method was then employed to automated detect the ARs observed by SDO/HMI during solar cycle 24 and the rising phase of solar cycle 25, and a statistical analyses of AR features was conducted. The unsupervised automated detection method is described in Section \ref{method}, followed by a description of its performance. The statistical results of ARs detected by aforementioned method are presented in Section~\ref{results} and compared with the published results in Section~\ref{discussion&conclusion}, where we also draw several conclusions.

\begin{figure}[ht!]
	\centering
	\includegraphics[width=\linewidth]{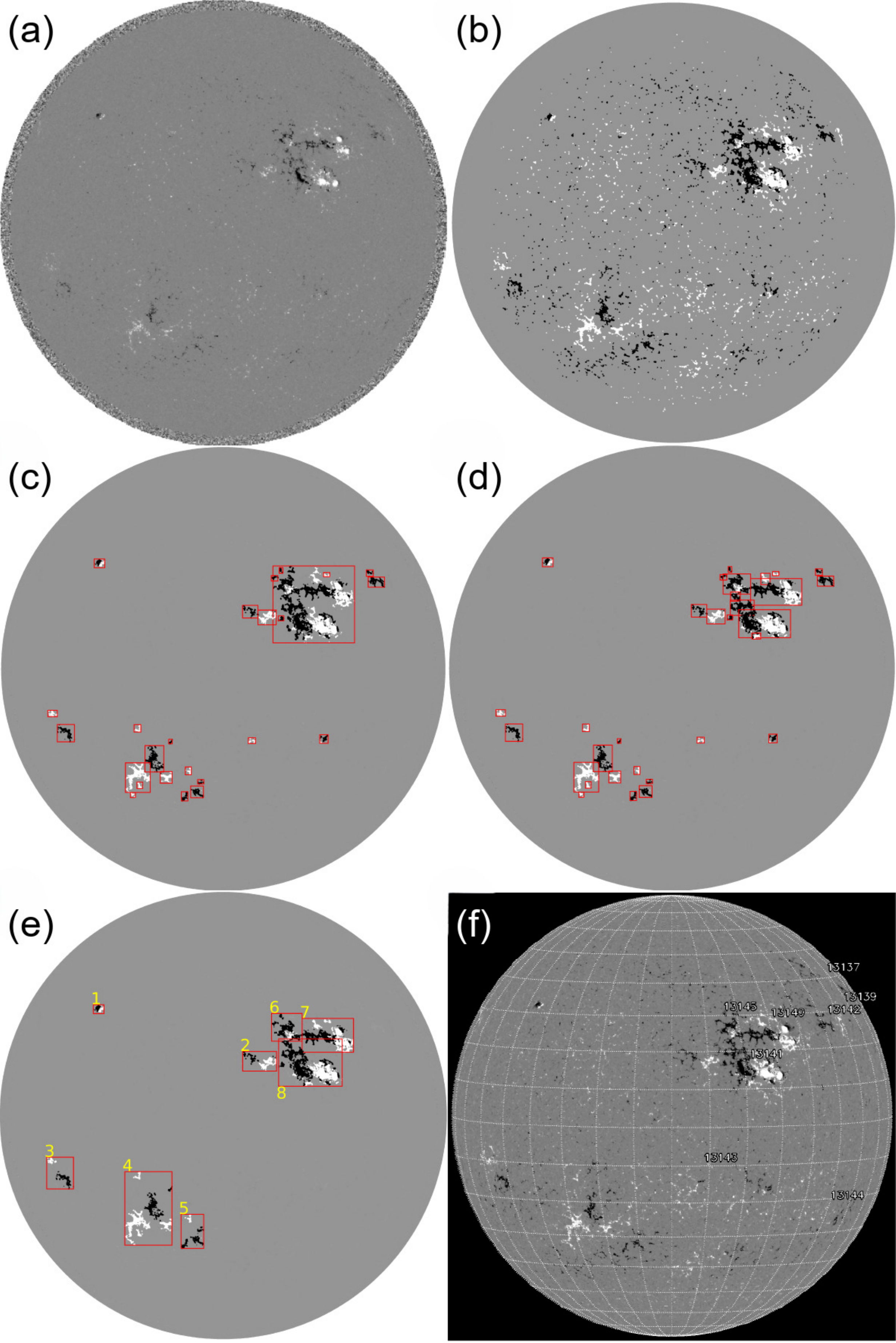}
	\caption{An observation taken on November 12, 2022 as an example to show the processing modules of the automated detection system. (a) The original observation obtained by SDO/HMI. (b) The magnetogram after threshold segmentation. (c) The detected result after the first (global) DBSCAN clustering, with each red box representing a cluster. (d) The detected result after the second (reformative) DBSCAN clustering. (e) The final detected AR results after filtering, merging, and labeling the clusters obtained from DBSCAN$^2$. Each red box with a number represents a bounding rectangle of the identified solar AR. (f) NOAA-identified solar active regions in the same day obtained from the SolarMonitor website (\url{https://solarmonitor.org}) for comparison.
	}
	\label{fig1}
\end{figure}

\section{Method} \label{method}

\subsection{Data Acquisition}\label{data}
The present study employs the line-of-sight (LOS) full-disk magnetograms observed by SDO/HMI. The SDO/HMI dataset has been acquiring LOS magnetograms at a 45 s cadence, with a spatial resolution 1 arcsec (with each image possessing $4096 \times 4096$ pixels) since the SDO launch on 2010 February 11. The dataset under consideration is in the form of a single image selected per day, spanning the period from May 2010 to December 2023, and is acquired through the Joint Science Operations Center\footnote{\url{http://jsoc.stanford.edu/HMI/Magnetograms.html}}. 

\subsection{DSARD: DBSCAN-based Solar Active Regions Detection}\label{auto-detect}

An automated detection method for solar active regions observed in magnetograms was developed, and this method was named DSARD (DBSCAN-based Solar Active Regions Detection). The method is based on the Density-Based Spatial Clustering of Applications with Noise (DBSCAN; \citealt{Schubert2017}), a popular clustering algorithm used in data analysis and machine learning to identify arbitrarily shaped clusters in imaging data. A significant advantage of DBSCAN over other clustering methods is its ability to effectively handle noise and outliers by categorizing points based on their density, eliminating the need to predefine the number of clusters. In essence, DBSCAN can classify magnetic field regions into distinct active regions without requiring prior knowledge of their number.

The core principle of DBSCAN relies on two key parameters: the distance criterion $\epsilon$ and the minimum number of pixel points required for a region to be considered dense, denoted by $P_{min}$. Based on these parameters, pixel points are categorised into three types:
\begin{itemize}
  \item \textbf{Core Point:} A point with at least $P_{min}$ points within its $\epsilon$-neighborhood.
  \item \textbf{Border Point:} A point that has fewer than $P_{min}$ points within its $\epsilon$-neighborhood but is in the neighborhood of a core point.
  \item \textbf{Noise Point:} A point that is neither a core point nor a border point.
\end{itemize}

After categorizing the points, clustering begins by identifying all core points and grouping them into clusters based on their reachability through other core points. Border points are then assigned to the nearest cluster, while noise points remain unclassified. This approach excels at discovering clusters of varying shapes and sizes and is robust against noises, making it particularly suitable for astronomical applications where traditional methods may fail.

The DSARD algorithm conceptualizes solar active regions in magnetograms as concentrations of pixels with relatively higher magnetic field intensities. This premise guided the initial application of grayscale threshold segmentation, consistent with previous methods \citep{Turmon2002, Zhang2010, Turmon2010}. However, DSARD goes beyond simple thresholding by considering both pixel connectivity and density distribution, leading to the implementation of DBSCAN. It is acknowledged that the performance of DBSCAN heavily depends on the selection of its hyperparameters. To address this, a double iteration of DBSCAN is employed to filter out noise points and generate clusters of varying shapes, as detailed below. The DSARD method consists of the following four steps:
\begin{enumerate}[leftmargin=*]
  \item \textbf{Threshold-based Image Segmentation:} The analysis commences with the implementation of a thresholding technique on the image, thereby accentuating pixels exhibiting substantial magnetic field intensities. The $threshold$ is set to be 150 Gauss, a level that encompasses a substantial proportion of pertinent pixels, analogous to the threshold employed for SHARP data \citep{Bobra2014}. The outcome of this segmentation process is depicted in Figure~\ref{fig1}(b), wherein positive and negative magnetic fields are delineated by white and black patches, respectively. However, the analysis reveals a pervasive distribution of noise across the solar disk.
  
  \item \textbf{First global DBSCAN:} In the first DBSCAN, a smaller value is assigned to $\epsilon$ and $P_{min_1}$ in order to filter out noise points remaining after threshold segmentation, thus resulting in the formation of preliminary clusters. It is evident that some clusters are larger than the true clusters due to the initial broad parameter settings, as illustrated in Figure~\ref{fig1}(c). The relatively larger cluster on the northwest side, which includes several active regions, is clearly visible. A second DBSCAN is employed to address this issue.
  
  \item \textbf{Second reformative DBSCAN:} In the subsequent DBSCAN processing, clusters that exceed a predefined size threshold, designated as $maxSize$, are designated as excessively large. In practice, $maxSize$ was set to $3000$ Mm$^2$. Consequently, the second DBSCAN was implemented with larger values for $P_{min_2}$. This step facilitates the segmentation the oversized clusters into more precise size. As Demonstrated in Figure~\ref{fig1}(d), the substantial cluster observed on the northwestern region of the solar disk in Figure~\ref{fig1}(c) has been meticulously divided into multiple smaller clusters.
  
   \item \textbf{Integration:} Following the procedure described above, relatively scattered clusters are obtained. As illustrated in Figure~\ref{fig1}(d), some clusters are found to be relatively small and/or to comprise unipolar regions. These regions should be integrated to form a complete active region. Initially, a parameter $ratio$ is set and the clusters are divided into three categories based on it. The definition of a ``positive cluster" is such that the number of points with positive magnetic polarity in the cluster exceeds the number of points with negative polarity by the factor of $ratio$. The definition of a ``negative cluster" is such that the number of points with negative magnetic polarity in the cluster exceeds the number of points with positive polarity by the factor of $ratio$. Clusters that do not fit these two categories are referred to as ``neutral clusters".
   
    Next, the distance between the closest points of any two clusters on the solar sphere is calculated. Due to computational constraints, the nearest points between two clusters on the image are identified, and their distance on the solar surface is approximated to be the distance between the two clusters. This distance is then multiplied by a coefficient $\alpha$, determined by the cluster categories (positive, negative, or neutral), to define a measurement of the distance between clusters. Notably, the distance coefficient between a positive and a negative cluster is relatively small. The values of $\alpha$ are summarized in Table~\ref{alpha}. 
   
    For each cluster, the nearest cluster is identified. If their distance is less than the minimum distance $minDistance$ (set to 50 Mm), they are merged. This process repeats until no further merging is possible. Additionally, neutral clusters with a pixel count below $minSize=75$ Mm$^2$ and any ``positive" or ``negative" clusters with a pixel count below $4\times minSize=300$ Mm$^2$ are discarded. The factor of four reflects the observational fact that these clusters often represent decayed active regions, where one polarity region is more diffuse or attenuated. After this process, clusters of appropriate size with reasonably matched positive and negative polarities are identified as active regions, as illustrated in Figure~\ref{fig1}(e).
\end{enumerate}

\begin{table}[ht]
\centering
\caption{Coefficient \(\alpha\) based on cluster attributes.}
\begin{tabular}{|c|c|c|c|}
\hline
$\alpha$ & \textbf{Positive} & \textbf{Negative} & \textbf{Neutral} \\ \hline
\textbf{Positive} & 0.5 & 0.5 & 1 \\ \hline
\textbf{Negative} & 0.5 & 0.5 & 1 \\ \hline
\textbf{Neutral} & 1 & 1 & 2 \\
\hline
\end{tabular}\label{alpha}
\end{table}

\begin{table}[h!]
  \begin{center}
    \caption{Parameters in our method.}
    \begin{tabular}{l|c} 
      \hline
      \textbf{Parameter} & \textbf{value} \\
      \hline
      $threshold$ & 150 Gauss \\
      $\epsilon$ & 30 pixels\\
      $P_{min_1}$ & 200 pixels\\
      $P_{min_2}$ & 500 pixels\\
      $maxSize$ & 3000 Mm$^2$\\
      $ratio$ & 10\\
      $minDistance$ & 50 Mm\\
      $minSize$ & 70 Mm$^2$\\
    \hline
    \end{tabular}\label{par}
  \end{center}
\end{table}

The advanced stages of segmentation and integration not only enhance the robustness of the algorithm to parameter variations but also effectively incorporate the morphological features of solar active regions into the computational framework, leveraging the magnetic intensity information from SDO/HMI magnetograms. The detailed parameters are summarized in Table~\ref{par}. For comparison, an example image with active regions identified by NOAA on the same day is shown in Figure~\ref{fig1}(f). In addition to NOAA ARs 13140, 13141, and 13145, our method identified five additional ARs, labeled as No. 1 to 5. NOAA ARs 13142, 13143, and 13144 were excluded due to their weak and diffuse nature. Besides, although our method accounts for the projection effects, it still fails if the active regions are near the solar limb, such as NOAA ARs 13139, 13142, and 13144.

\begin{figure}[ht!]
	\centering
	\includegraphics[width=\linewidth]{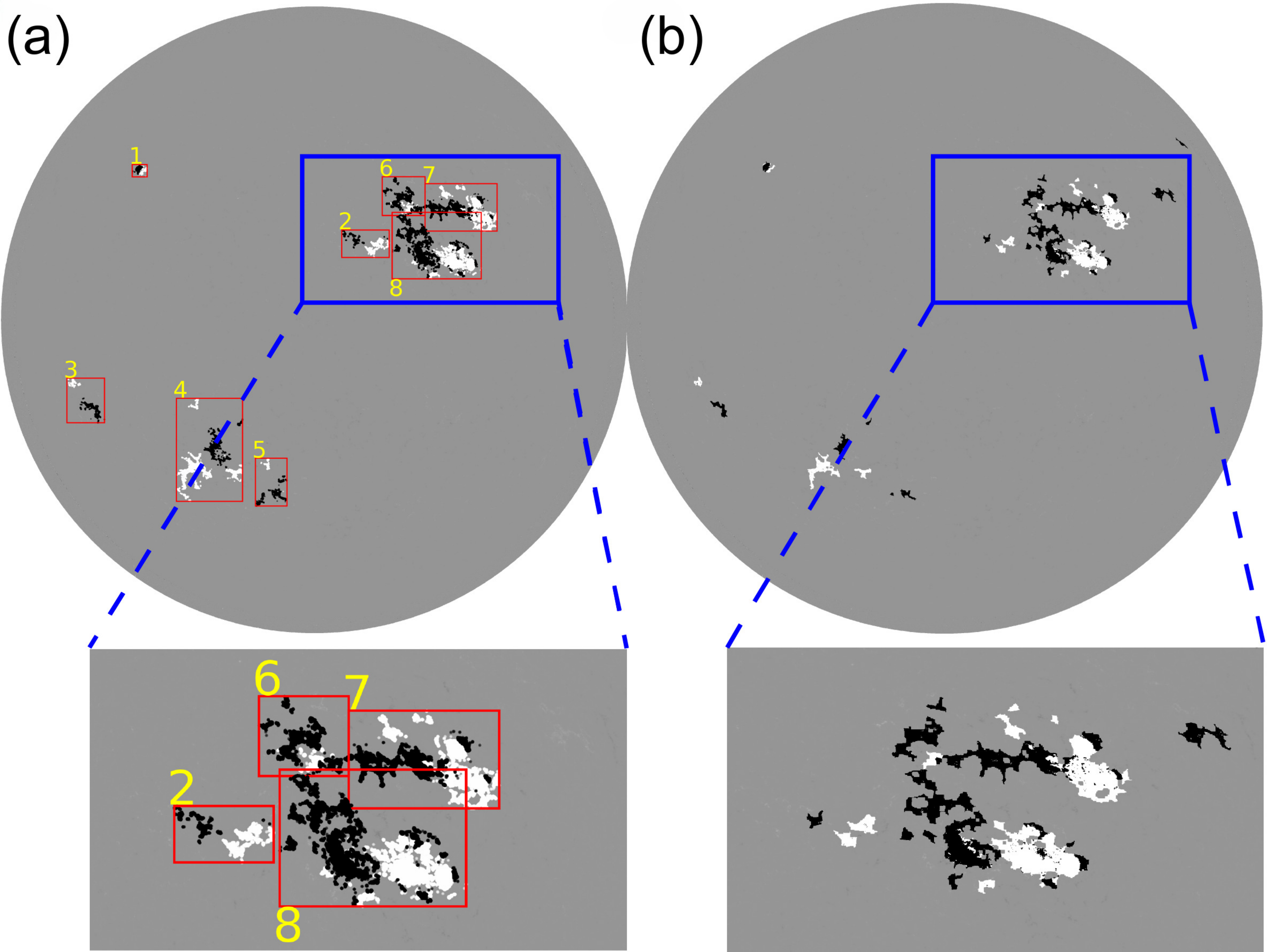}
	\caption{An example to compare the detected AR results by the present method and the classical image processing method. (a) The results detected by the present method are shown in the top panel, which are equivalent to those illustrated in Figure~\ref{fig1}(e). The lower panel displays an enlarged section of the blue box region. (b) The results detected by the classical image processing method.
	}
	\label{fig2}
\end{figure}

\begin{figure}[ht!]
    \centering
    \includegraphics[width=0.48\textwidth]{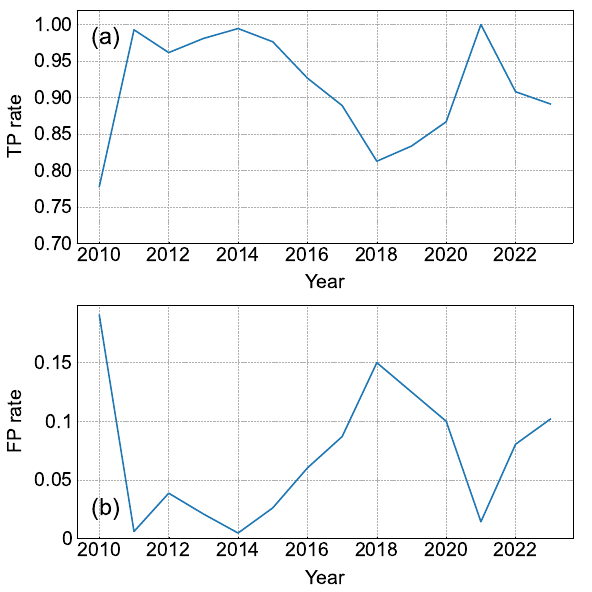} 
    \caption{Variation of true positive rate $R_{TP}$ (a) and false positive rate $R_{FP}$ (b) from 2010 to 2023. }
    \label{fig3}
\end{figure}

\subsection{Performance}\label{perform}

The effectiveness of the above-mentioned identification algorithm has been rigorously validated, showcasing notable advantages in two key areas. First, the primary strength of the DSARD method lies in its exceptional sensitivity to the data density, enabling the detection of smaller or more diffuse ARs that might be overlooked by other algorithms or even manual identification methods. This capability is demonstrated in Figures~\ref{fig1}(e) and \ref{fig1}(f), where the DSARD method identified ARs labeled No. 1 and 3 to 5 in the northeast and southeast quarters of the solar disk—regions not marked by NOAA. Second, the sophisticated integration of connectivity and density considerations in our algorithm allow for a more refined segmentation of adjacent ARs. This approach differs significantly from classical image processing techniques, such as region-growing methods, which tend to indiscriminately group all connected areas. Figures~\ref{fig2}(a) and (b) provide a comparison between the identification results of the DSARD method and those of the region-growing method. The region-growing method fails to capture some nuances of ARs and incorrectly merges distinct ARs due to its inherent limitations. While morphological open operations are often used to disconnect linked ARs, they result in smaller detected AR areas, as shown in Figure~\ref{fig2}(b). In contrast, the DSARD method effectively distinguishes complex ARs while preserving their boundaries, as illustrated in the enlarged panel of Figure~\ref{fig2}(a).

Although the algorithm calculates distances at the pixel scale, resulting in slightly longer processing times compared to conventional methods, the processing speed remains acceptable, i.e., approximately 30 seconds per $4096\times 4096$ pixel image on a standard personal computer. The DSARD method and the detected AR features are available on GitHub\footnote{\url{https://github.com/Chenruishuo/DSARD}}. Additionally, copies of the method and data are archived on Zenodo\footnote{\url{https://doi.org/10.5281/zenodo.14222292}}.

To quantitatively evaluate the performance of our method, the NOAA catalog\footnote{ \url{http://solarcyclescience.com/activeregions.html}.} serves as the ``ground truth”. We processed 4984 fulldisk magnetograms from May 2010 to December 2023, detecting 25,178 ARs. The detection process is applied to a full-disk magnetogram per day, meaning the same active region (AR) may be detected multiple times. To address the issue of duplicated detections and minimize the projection effects, we focus on a longitudinal range of $\pm 6^{\circ}$ from the central meridian of the solar disk. Within the specified range, the final number of ARs is 2863. Detailed AR counts can be found in Table~\ref{results}. For comparison, we only consider the NOAA-labeled ARs within the specified longitudinal range. The true positive rate $R_{TP}$ and the false positive rate $R_{FP}$ are then defined, following the approach of \citet{Zhang2010}:
\begin{eqnarray}
	R_{TP} = \frac{N_{TP}}{N_{TP} + N_{FN}}=\frac{N_{TP}}{N_{NOAA}}\,,
\end{eqnarray}
\begin{eqnarray}
	R_{FP} = \frac{N_{FP}}{N_{FP} + N_{TN}}=\frac{N_{FP}}{N_{FP}+N_{NOAA}}\,,
\end{eqnarray}
where $N_{TP}$ represents the number of true positives, defined as cases where the center locations reported by NOAA fall within the minimum bounding rectangle of the detected ARs; $N_{FP}$ denotes the number of false positives, where detected ARs are not recorded in the NOAA catalog; $N_{FN}$ refers to the number of false negatives, indicating ARs in the NOAA catalog that the automated method fails to detect; $N_{TN}$, the number of true negatives, cannot be determined since there is no concept of a ``negative AR event". As shown in Figure~\ref{fig3}, the true positive rate $R_{TP}$ and false positive rate $R_{FP}$ vary from 2010 to 2023, as indicated by the panels (a) and (b), respectively. The average $R_{TP}$ and $R_{FP}$ are $91.8\%$ and $7.2\%$, respectively. It is notable that the relatively low $R_{TP}$ is evident in the years 2010 and around 2018, both of which are due to the scarcity of ARs near solar minimum.

\begin{figure}[ht!]
    \centering
    \includegraphics[width=0.48\textwidth]{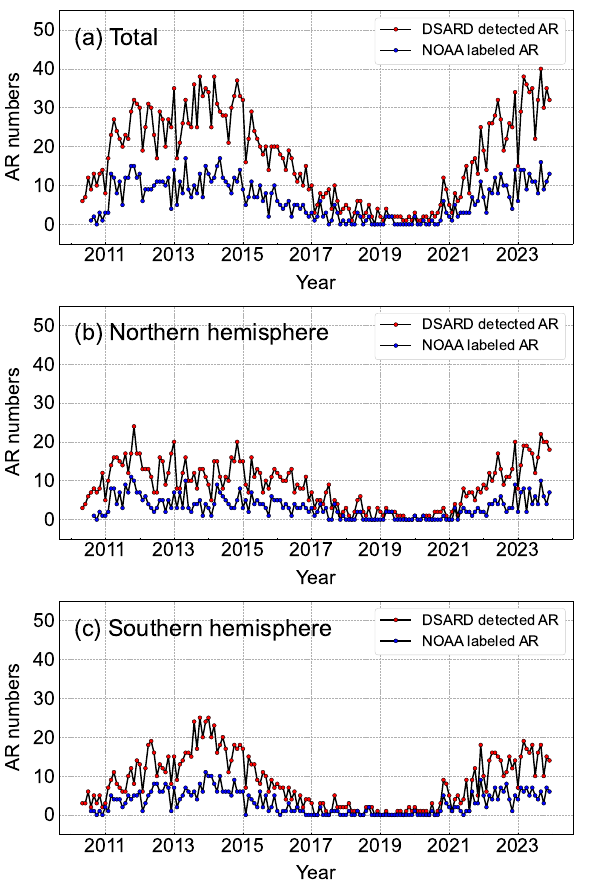}  
    \caption{Distribution of the AR number detected by our method and NOAA catalog from 2010 to 2023. (a) Distribution of the total AR number from 2010 to 2023. Each dot represents the sum of the AR number per month. Panels(b) and (c) are similar to panel (a) but for the number of ARs in the northern and southern hemispheres, respectively.}
    \label{fig4}
\end{figure}

The yearly variation of the numbers of ARs detected by the proposed method and from the NOAA catalog during 2010 -- 2023 is illustrated in Figure~\ref{fig4}. While the distribution of AR numbers detected by the proposed method follows a similar trend to that of the NOAA catalog, the former consistently identifies more ARs. The discrepancies in AR counts are more pronounced in the southern hemisphere compared to the northern hemisphere, particularly around year 2014. This increased detection of ARs can be attributed to the enhanced sensitivity of the DBSCAN-based method to the data density, enabling it to identify smaller or more diffuse ARs that may be overlooked by other methods.

\begin{table*}  \scriptsize
\centering
\caption{Automatically detected AR numbers from 2010 to 2023.} \label{results}
\begin{tabular}{ccccccccc} \\
\hline \hline

                        &               &               &			&           &Detected   &AR     &Numbers           &  \\
Time interval          	&Solar Cycle    &Data numbers   &	        &Full Disk  &			&	        &$\pm 6^{\circ}$ &   \\
      		            &               &               &Total      &North      &South      &Total      &North      &South   \\
\hline
2010 May -- 2019 Dec    &24             &$3,523$        & $17,523$ 	&$9,267$   &$8,256$   &$1,991$      &$1,037$   &$954$\\
\hline
2020 Jan -- 2023 Dec    &25             & $1,461$       & $7,655$  &$3,798$   &$3,875$   &$872$      &$430$   &$442$\\
\hline
\end{tabular}
\end{table*}

\begin{figure}[ht!]
    \centering
    \includegraphics[width=0.48\textwidth]{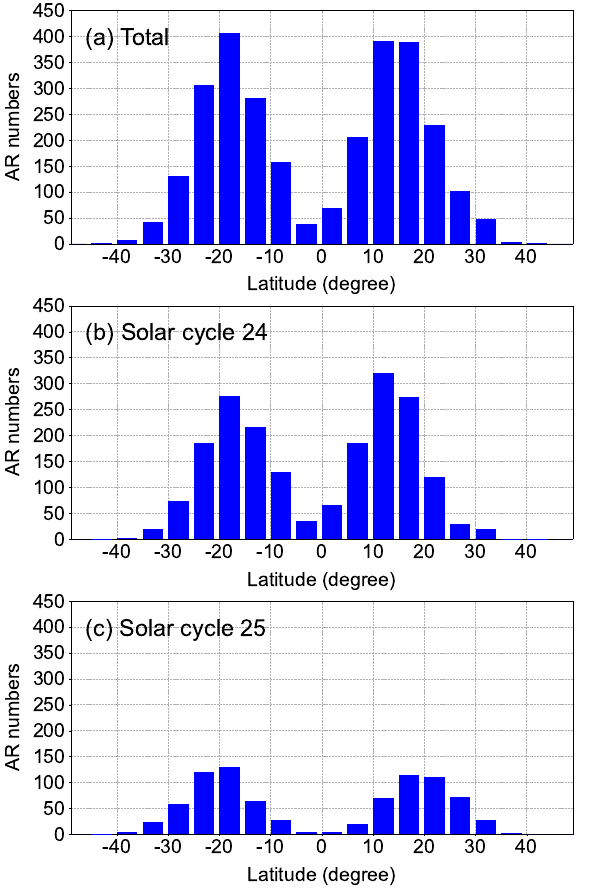}
    \caption{Histograms of active regions showing their distribution with latitude. (a) Total events; (b) Active regions in solar cycle 24; (c) Active regions in the rising phase of solar cycle 25.}
    %\caption{Distribution of the number of solar active regions with respect to latitude. (a) Histogram of the distribution of the number of solar active regions with respect to latitude during the solar cycle 24. The horizontal axis represents solar latitude and the vertical axis represents the number of solar active regions within each latitude interval. (b) Histogram of the distribution of the number of solar active regions with respect to latitude during the rising phase of solar cycle 25.(c) Histogram of the distribution of the overall number of solar active regions with respect to latitude from May 2010 to December 2023.}
    \label{fig5}
\end{figure}

\section{Results}\label{results}

As outlined in Section~\ref{perform}, the method employed in this study processes the SDO/HMI line-of-sight magnetograms on a daily basis from May 2010 to December 2023, covering most of solar cycle 24 and the rising phase of solar cycle 25. To avoid multiple detections and minimize projection effects, a longitudinal range limit of $\pm 6^{\circ}$ from the central meridian of the solar disk is applied. Within this range, 1991 ARs were detected during solar cycle 24 and 872 ARs during the rising phase of solar cycle 25. The detailed counts of the detected ARs are summarized in Table~\ref{results}. In addition to generating butterfly diagrams and analyzing drift velocities, the geometric center (location), area, magnetic flux, and tilt angles of bipolar ARs are calculated for different years and latitude bands. The statistical results are presented in detail in the following subsections.

\subsection{Distributions of the Number of Active Regions}

As mentioned in Section~\ref{perform}, Figure~\ref{fig4} reveals a strong correlation between the yearly variation of AR numbers detected by our method and those listed in the NOAA catalog from 2010 to 2023. Both datasets indicate the presence of two peaks in AR numbers during solar cycle 24, with the first occurring around 2012 and the second, more pronounced peak around 2014. Notably, the first peak is dominated by ARs in the northern hemisphere, while the second peak is dominated by those in the southern hemisphere, as illustrated in Figures~\ref{fig4}(b) and (c). During the solar maximum of cycle 24, the monthly AR count exceeded 20, but it declined to less than 5 or even 0 per month during the declining phase from 2018 to 2020. In contrast, during the rising phase of solar cycle 25, the AR numbers increased rapidly and are significantly higher compared to those of solar cycle 24, suggesting that the strength of the rising phase of solar cycle 25 may surpass that of solar cycle 24.

The latitudinal distribution of ARs is illustrated in Figure~\ref{fig5}, showing a bimodal distribution with ARs primarily concentrated in the latitude band of  $10^{\circ}$--$20^{\circ}$. Few ARs are observed at latitudes higher than $40^{\circ}$. The analysis indicates that during solar cycle 24, the peak in the southern hemisphere occurs in the latitude band $15^{\circ}$--$20^{\circ}$, while in the northern hemisphere, it is in the $10^{\circ}$--$15^{\circ}$ band. In contrast, during the rising phase of solar cycle 25, the peaks appear in the $15^{\circ}$--$20^{\circ}$ band in both hemispheres. Notably, during solar cycle 24, there are more active regions in the northern hemisphere than in the southern hemisphere, and it is reversed during the rising phase of solar cycle 25.

\begin{figure*}[hbt!]
    \centering
    \includegraphics[width=\textwidth]{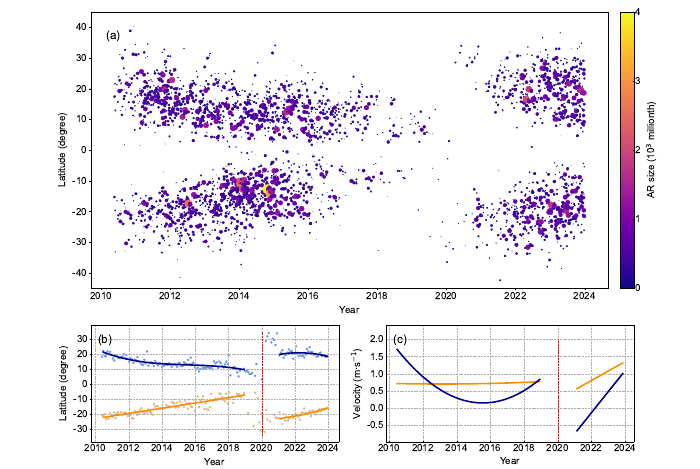}

    \caption{Butterfly diagram and latitudinal migration of ARs during the solar cycle 24 and the rising phase of solar cycle 25. (a) Butterfly diagram of active regions from 2010 to 2023. Each dot represents a single AR, with the size and color of the dot corresponding to the respective area of the active region. (b) Temporal evolution of the monthly average latitude of active regions in the northern (blue dots) and southern (orange dots) hemispheres is illustrated, with the blue and orange lines representing the cubic polynomial fitted lines of the monthly average latitude of active regions. The dash vertical line indicates the beginning of solar cycle 25. (c) Fitted drift velocity variations in the northern (blue line) and southern (orange line) hemispheres, respectively.}
    \label{fig6}
\end{figure*}

\subsection{Butterfly Diagram and Drift Velocity variations of Active Regions}

The latitudinal distribution of ARs over a solar cycle forms a ``butterfly diagram", a pattern known as ``Sp{\"o}rer’s Law of Zones” \citep{Hathaway2010}. This latitudinal migration toward the equator reflects the meridional flow associated with the solar dynamo \citep{Choudhuri2021}. The butterfly diagram for ARs during solar cycle 24 and the rising phase of solar cycle 25 is presented in Figure~\ref{fig6}. Most ARs are located below $40^{\circ}$, with only a few exceptions. The relatively large ARs are predominantly found in the low-latitude band of $10^{\circ}$--$15^{\circ}$  in both hemispheres. During solar cycle 24, the ARs in the northern hemisphere appear for a longer time compared to those in the southern hemisphere. Specifically, ARs in the northern hemisphere emerged from 2010 and persisted until the end of solar cycle 24 in 2020, gradually decreasing in both area and number each year. In contrast, ARs in the southern hemisphere increased in number and area from 2010, peaked around 2014, and then declined rapidly by 2016, with only a few appearing after 2018. Relatively large ARs began to appear around 2012 in the northern hemisphere, while in the southern hemisphere, they emerged since late 2012 and became particularly large and numerous in 2014 and 2015. This suggests that ARs in the southern hemisphere were stronger than those in the northern hemisphere during these years. During the rising phase of solar cycle 25, ARs began to emerge more intensely than in solar cycle 24. A notable feature of this phase is the earlier appearance and larger size of ARs in the southern hemisphere compared to the northern hemisphere. Additionally, there was a gradual increase in both the number and area of ARs from late 2021 to 2023.

To conduct a quantitative analysis of AR migration during solar cycle 24 and the rising phase of solar cycle 25, the average latitude of ARs is calculated at 30-day intervals. These average latitudes are then fitted using a cubic polynomial. Figure~\ref{fig6}(b) illustrates the evolution of the monthly average latitudes of ARs during solar cycles 24 and 25, with blue and orange dots representing the respective cycles. The blue and orange lines correspond to the cubic polynomial fits of the monthly average latitudes, while the dashed vertical line marks the start of solar cycle 25. Due to the dispersed latitude distribution around 2020, data from this period are excluded from the analysis. The monthly average latitudes of ARs in both hemispheres begin at approximately 
$20^{\circ}$, which are lower than the latitudes shown in the butterfly diagram of all ARs. This discrepancy arises because the majority of ARs are located below $25^{\circ}$, as demonstrated by the histograms of the latitudinal distribution of AR numbers in Figure~\ref{fig5}.

As shown in Figure~\ref{fig6}(c), the derived drift velocity evolutions in the northern and southern hemispheres are represented by blue and orange curves, respectively. During solar cycle 24, ARs in both hemispheres migrated toward the equator, displaying significant asymmetry. ARs in the southern hemisphere exhibited relatively stable drift velocity variations. A decline in drift velocities was observed during the rising phase around 2014, followed by a modest increase toward the end of solar cycle 24, with maximum and minimum velocities of 0.8 m s$^{-1}$ and 0.7 m s$^{-1}$, respectively. In contrast, the drift velocities of ARs in the northern hemisphere showed greater variability compared to those in the southern hemisphere. A decline from 1.5 m s$^{-1}$ to 0.1 m s$^{-1}$ was observed in late 2015, followed by an increase to 0.8 m s$^{-1}$ by the end of solar cycle 24 in 2019. The rising phase of solar cycle 25 also manifested clear asymmetries in drift velocities between the two hemispheres. In the northern hemisphere, a deceleration of velocities toward relatively high latitudes occurred from 2021 to 2022, followed by an acceleration toward the equator, reaching 1.0 m s $^{-1}$ in 2023. In contrast, the southern hemisphere demonstrated a consistent acceleration toward the equator, with velocities increasing from 0.5 m s$^{-1}$ to 1.2 m s$^{-1}$.

\begin{figure}
    \centering
    \includegraphics[width=0.5\textwidth]{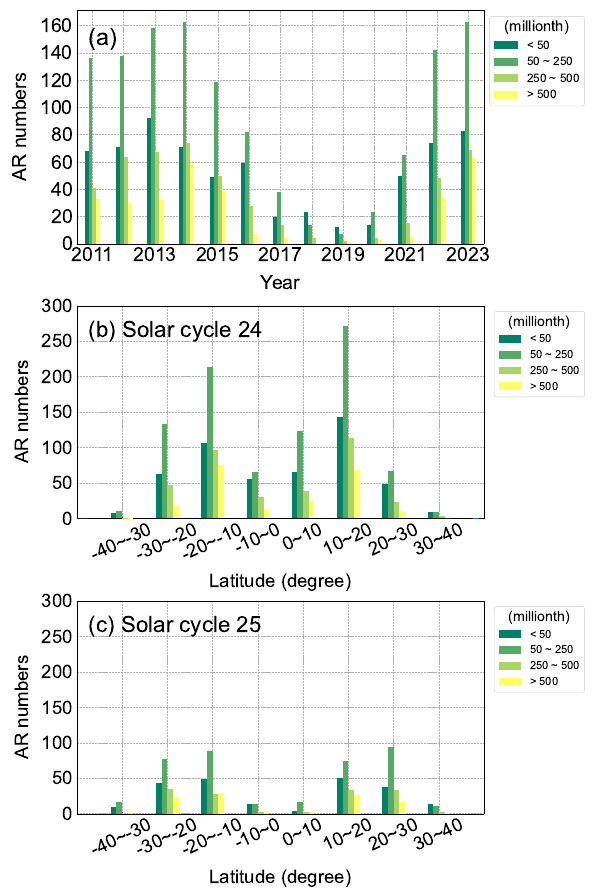}
    
    \caption{Distribution of AR numbers in different area ranges with respect to latitude and year from 2010 to 2023.(a) Distribution of AR numbers in different area ranges with respect to year. (b)Distribution of AR numbers in different area ranges with respect to latitude during solar cycle 24. Panel (c) is similar to (b) but for that during the rising phase of solar cycle 25.}
    \label{fig7}
\end{figure}

\subsection{Distributions of Active Region Areas}

The area of ARs is measured in millionths, defined as millionths of the total solar surface area. AR areas are categorized into four groups based on their size: 
$<$50 millionths, 50--250 millionths, 250--500 millionths, and $>$500 millionths. The proportions of these groups are approximately $1:2:4:2$. Figure~\ref{fig7}(a) illustrates the yearly distributions of AR numbers of these four area groups. In general, the number of ARs increased from the beginning of the solar cycle to the solar maximum and then decreased toward the end of solar cycle 24. The peak in AR numbers occurred in 2014, except for the group with areas $<$50 millionths, which shows two peaks in 2013 and 2016 during solar cycle 24. During the rising phase of solar cycle 25, the number of ARs in all four groups increased steadily from 2020 to 2023.

As illustrated in Figures~\ref{fig7}(b) and \ref{fig7}(c), the latitudinal distribution of AR numbers of the four area groups during solar cycle 24 and the rising phase of solar cycle 25 exhibits a bimodal pattern. During solar cycle 24, the peaks are observed in the latitude band of $10^{\circ}$--$20^{\circ}$ in both hemispheres. Within the latitude band of $0^{\circ}$--$20^{\circ}$, the AR numbers in the four area groups are higher in the northern hemisphere compared to the southern hemisphere. Conversely, in the latitude band of $20^{\circ}$--$30^{\circ}$, the southern hemisphere shows higher AR numbers. In contrast, during the rising phase of solar cycle 25, no significant hemispheric asymmetry is observed. The peaks in AR numbers predominantly occur in the $10^{\circ}$--$20^{\circ}$ latitude band, with the exception of ARs in the northern hemisphere with areas of 50--250 millionths and those in the southern hemisphere with areas of 250--500 millionths. Notably, the AR numbers in the $30^{\circ}$--$40^{\circ}$ latitude band during the rising phase of solar cycle 25 are comparatively higher than those in solar cycle 24, suggesting a broader distribution of ARs at relatively higher latitudes in solar cycle 25.

\begin{figure*}[ht!] 
    \centering
    \includegraphics[width=0.9\textwidth]{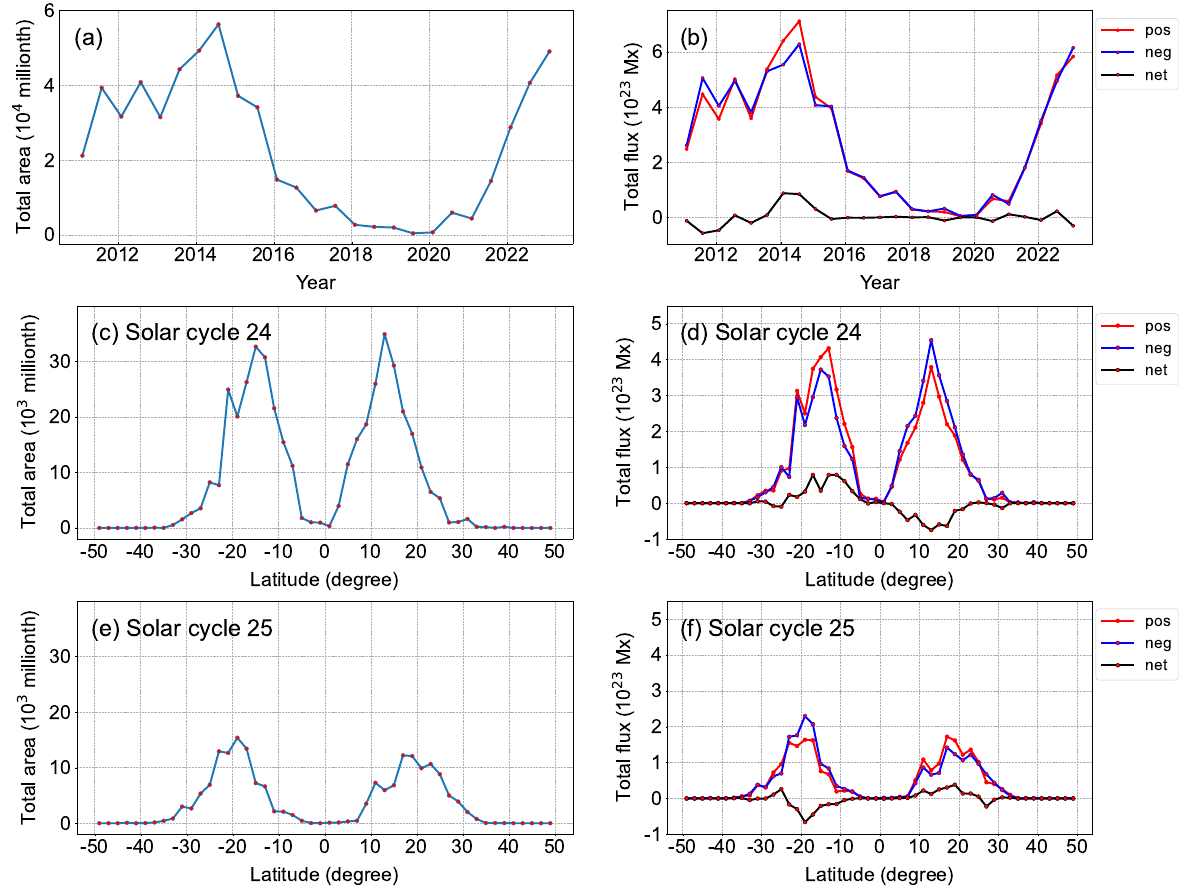}
    \caption{Distribution of cumulative area and magnetic flux with respect to latitudes and years.}
    \label{fig8}
\end{figure*}

As shown in Figures~\ref{fig8}(a), \ref{fig8}(c), and \ref{fig8}(e), the yearly and latitudinal distribution of cumulative active region (AR) areas reveals a prominent peak in the late part of 2014, consistent with the AR number trends shown in Figure~\ref{fig7}(a). The cumulative AR areas exhibit a more rapid increase during the rising phase of solar cycle 25 compared to solar cycle 24. Additionally, the cumulative AR areas across different latitude bands display a bimodal distribution, as illustrated in Figures~\ref{fig8}(c) and \ref{fig8}(e). The largest cumulative areas are observed in the latitude band of $10^{\circ}$ -- $20^{\circ}$ in both hemispheres. Notably, the latitudes of the peaks during the rising phase of solar cycle 25 are higher than those observed in solar cycle 24.

\begin{figure*}[ht!] 
    \centering
   \includegraphics[width=\textwidth]{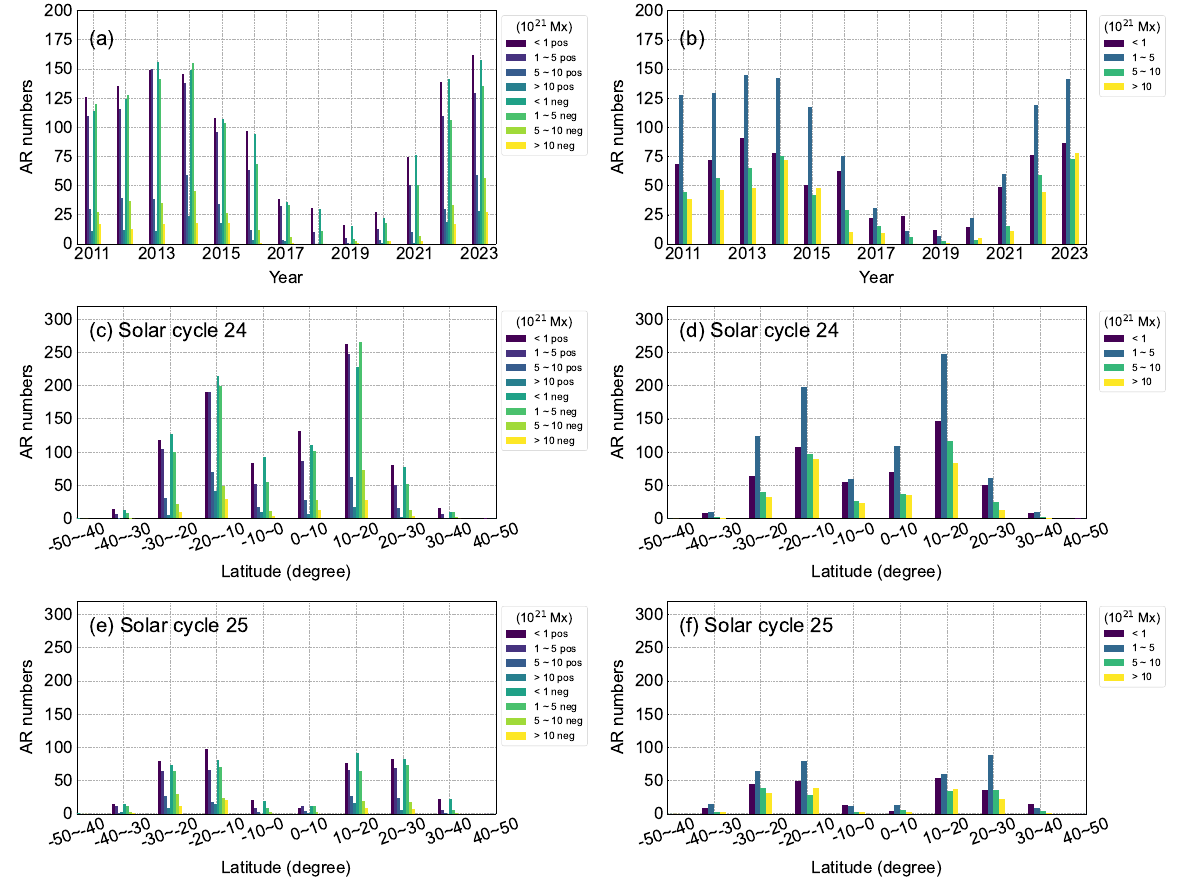}

    \caption{Similar to Figure~\ref{fig7} but for the magnetic flux.}
    \label{fig9}
\end{figure*}

\subsection{Distributions of Active Region Magnetic Flux}

The magnetic flux is calculated by summing the magnetic fluxes threading through the pixels within each AR. This process yields the positive, negative, and total unsigned fluxes for each individual AR. Similar to the classification of AR areas, the magnetic fluxes of ARs are categorized into intensity-based groups: $< 10^{21}$ Mx, 1--5 $\times10^{21}$ Mx, 5--10 $\times10^{21}$ Mx, and $>10^{22}$ Mx. The yearly distributions of the positive, negative, and total unsigned fluxes for these four magnetic flux groups are shown in Figures~\ref{fig9}(a) and (b). Most ARs with positive and negative fluxes are below $5\times10^{21}$ Mx , while the total unsigned fluxes are predominantly below $10^{22}$ Mx. The peak flux for ARs with fluxes less than $10^{21}$ Mx occurs in 2013, whereas ARs with larger fluxes (1--5 $\times10^{21}$ Mx) peak one year later, in 2014, during solar cycle 24. For ARs with fluxes within 5--10 $\times10^{21}$ Mx, two peaks are observed in 2012 and 2014, though the first peak is less pronounced. ARs with magnetic fluxes exceeding $10^{22}$ Mx appear exclusively during the rising phase and extend into the years around the solar maximum of solar cycle 24. Notably, ARs of all four magnetic flux groups exhibit an increasing trend during the rising phase of solar cycle 24.

Figures~\ref{fig9}(c)--(f) illustrate the latitudinal distribution of ARs with positive, negative, and total unsigned fluxes across four magnetic flux groups during solar cycle 24 and the rising phase of solar cycle 25. The distribution of ARs in these flux groups is consistent with the distribution of AR areas. The latitudinal distribution of AR magnetic flux exhibits a bimodal pattern, with peaks in the latitude band of $10^{\circ}$--$20^{\circ}$ in both hemispheres. ARs with larger magnetic fluxes (exceeding $10^{22}$ Mx) are predominantly found at lower latitudes, below $30^{\circ}$. During solar cycle 24, ARs with magnetic fluxes below 10 $\times10^{21}$ Mx are more numerous in the northern hemisphere compared to the southern hemisphere. However, no significant hemispheric asymmetry is observed during the rising phase of solar cycle 25. Additionally, ARs with relatively small fluxes (less than 1 $\times10^{21}$ Mx) show a predominance of positive flux in the southern hemisphere and negative flux in the northern hemisphere.

Figures~\ref{fig8}(b), (d), and (f) display the yearly and latitudinal distribution of the cumulative magnetic flux of ARs. The largest cumulative values of positive, negative, and net magnetic fluxes are observed in 2014, consistent with the trends in AR numbers and AR areas during solar cycle 24. The similar variation trends for cumulative positive and negative magnetic fluxes, as well as cumulative areas, arise from their strong correlation. Notably, the cumulative areas increased more rapidly during the rising phase of solar cycle 25 compared to solar cycle 24. The cumulative positive and negative magnetic fluxes across different latitude bands also exhibit a bimodal distribution, as shown in Figures~\ref{fig8}(c) and (b), with peaks occurring in the latitude band of $10^{\circ}$ -- $20^{\circ}$ in both hemispheres. The net magnetic flux is negative from 2010 to 2013, becomes positive with relatively large values in 2014, and remains near zero during the decay phase of solar cycle 24. During the rising phase of solar cycle 25, the positive and negative magnetic fluxes show a rapid increase, while the net magnetic flux remains approximately zero. The latitudinal distribution of the cumulative net magnetic flux of ARs reveals that, during solar cycle 24, the net magnetic flux is positive in the southern hemisphere and negative in the northern hemisphere within the low-latitude bands of $0^{\circ}$ -- $30^{\circ}$. However, this pattern reverses during the rising phase of solar cycle 25, albeit with relatively weaker strength.

\begin{figure}[ht!]
    \centering
    \includegraphics[width=0.48\textwidth]{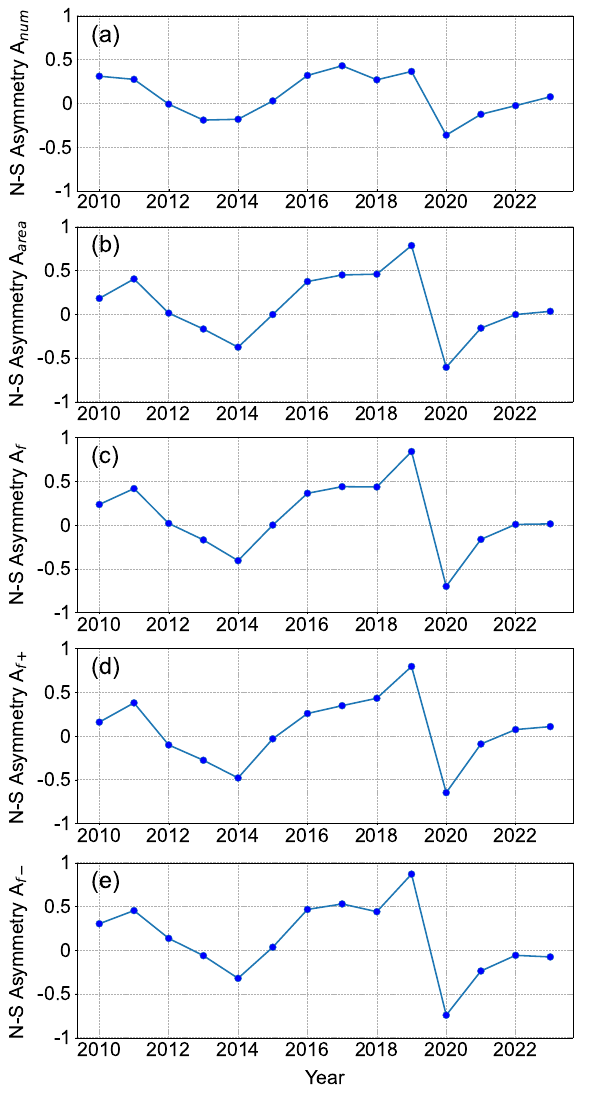}  
    \caption{Yearly asymmetries of the number (panel a), cumulative area (panel b), total flux (panel c), cumulative positive (panel d) flux and negative flux  (panel e) of all ARs.}
    \label{fig10}
\end{figure}

\subsection{N-S Asymmetry of Active Regions}

The N-S asymmetry is a typical characteristic of various solar features, which changes during each solar cycle \citep{Li2010, Hao2015, Chowdhury2019, Veronig2021, Chandra2022,Zhukova2024}. The N-S asymmetries of the numbers, areas, and magnetic flux of ARs are also investigated in this work. The N-S asymmetry index is defined as
	\begin{equation}
		A_{num}=\frac{N_n-N_s}{N_n+N_s},
	\end{equation}
	\begin{equation}
		A_{area} \ {\rm or }\ A_{f^{+}}\ {\rm or}\ A_{f^{-}}\ {\rm or}\ A_{f}=\frac{C_n-C_s}{C_n+C_s},
	\end{equation}
where $N$ and $C$ denote the numbers and the cumulative area or magnetic flux, while $n$ and $s$ represent the northern and southern hemispheres, respectively.  $A_{num}$ is the N-S asymmetry index for the AR number, $A_{area}$,  $A_{f^+}$, $A_{f^-}$, and $A_{f}$  are the N-S asymmetry indices for the normalized cumulative positive, negative, and total unsigned magnetic flux, respectively. If the sign of the N-S asymmetry index is positive/negative, it means that the northern/southern hemisphere is dominant. 

The yearly asymmetries of the AR number, normalized cumulative area, and fluxes are plotted in Figure~\ref{fig10}. The asymmetry indices for the AR number, cumulative area, and total unsigned magnetic flux indicate that the northern hemisphere is dominant for most of the time. The southern hemisphere becomes dominate during 2013--2014 and 2020--2021, and no asymmetry is discernable in 2012, 2015, and 2022. The asymmetry indices for the cumulative positive and negative fluxes exhibit similar trends, with slight deviations. Specifically, the northern hemisphere dominates in the years 2010--2011, 2016--2019, and 2022--2023, while the southern hemisphere dominates in 2012--2015 and 2020--2021. This suggests that the northern hemisphere is dominant for ARs with both positive and negative magnetic flux during the rising and decay phases of solar cycle 24, whereas the southern hemisphere dominates around the solar maximum. During the rising phase of solar cycle 25, the cumulative positive magnetic flux weakly dominates in the northern hemisphere, and the cumulative negative magnetic flux weakly dominates in the southern hemisphere.

\subsection{Distributions of the Tilt Angle of Bipolar Active Regions}

\begin{figure*}[ht!]
    \centering
    \includegraphics[width=0.8\textwidth]{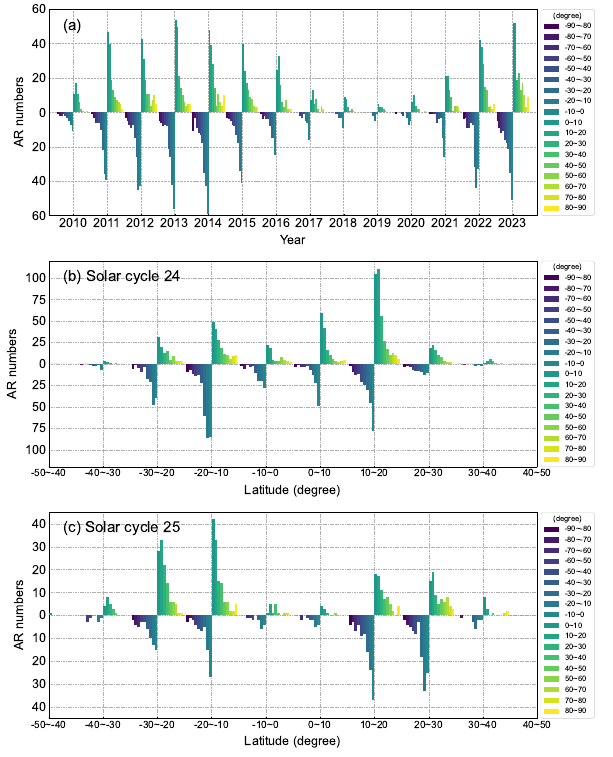}
    \caption{Distribution of the tilt angle of magnetic dipole in ARs with respect to years and latitude from 2010 to 2023. (a) Distribution of the tilt angle of magnetic dipole in ARs with respect to years. (b) Distribution of the tilt angle of magnetic dipole in ARs with respect to latitude during solar cycle 24. Panel (c) is similar to (b) but for those during the rising phase of solar cycle 25.
    }
    \label{fig11}
\end{figure*}

\begin{figure*}[ht!]
	\centering
	\includegraphics[width=\linewidth]{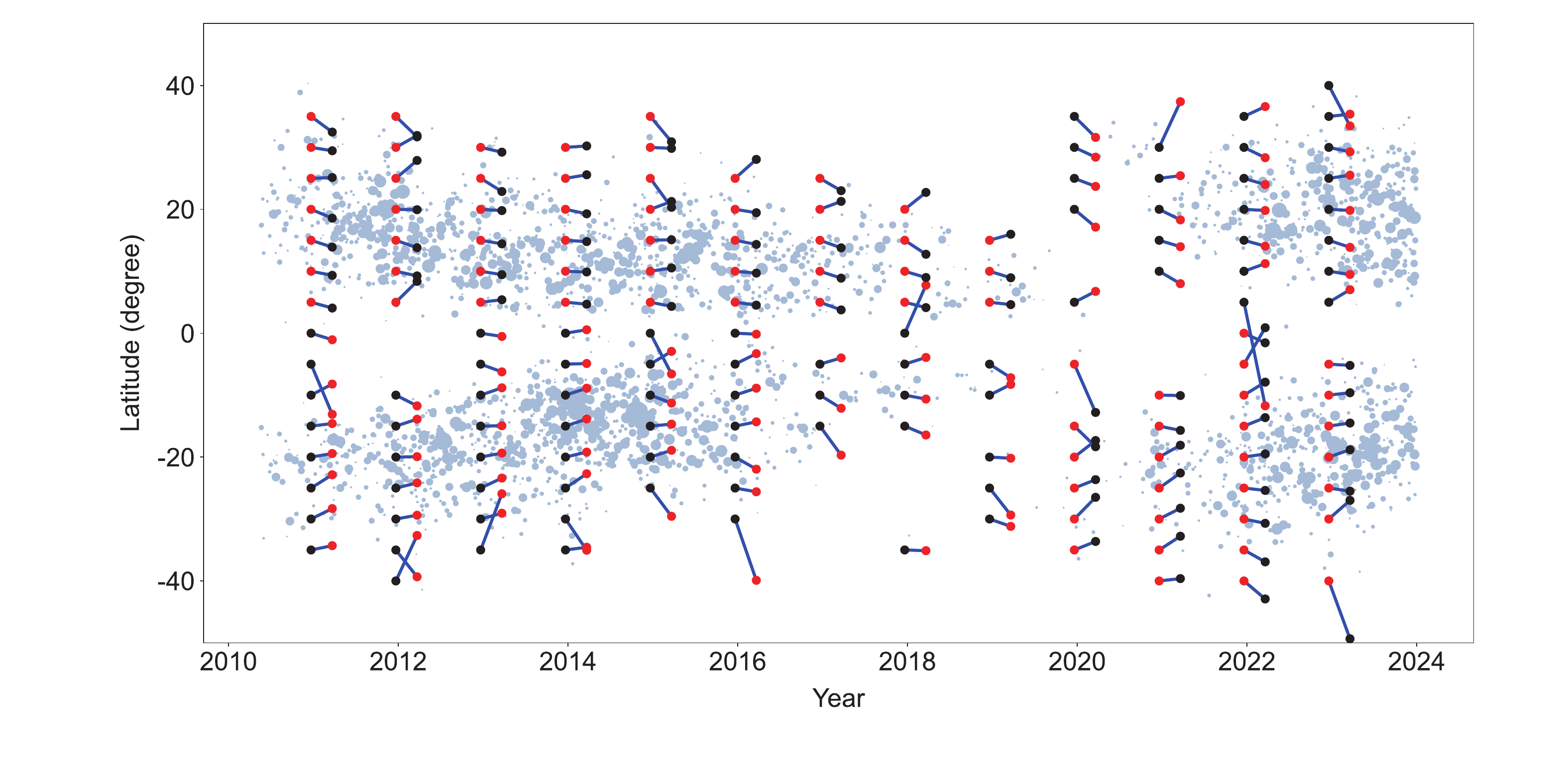}
    \caption{Distribution of average relative latitude of magnetic bipolar ARs with respect to years. The butterfly diagram is represented by light blue dots. The average latitudes of positive and negative polarities are indicated by red and black dots, respectively. The lines connecting the bipoles illustrate the tilt angle at specific latitude and time. To present the difference in tilt angles more clearly, they are multiplied by a factor of 8 to enlarge the angle differences. 
	}
	\label{fig12}
\end{figure*}

\begin{figure*}[ht!] 
    \centering
    \includegraphics[width=0.8\textwidth]{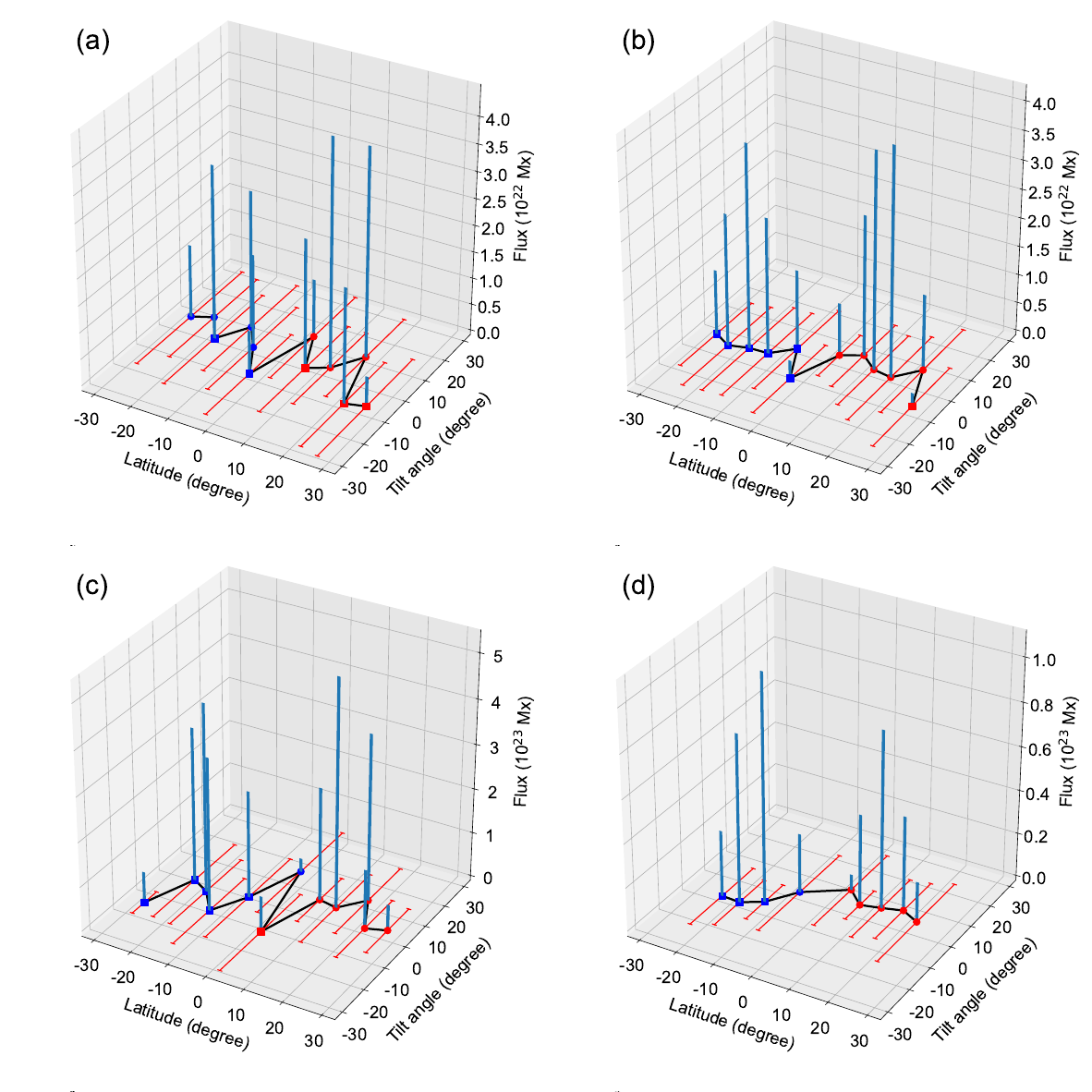}
    \caption{Latitudinal distribution of average ARs tilt angles in every $5^{\circ}$ interval during solar cycle 24. Panels (a)-- (d) are the latitudinal distribution of average ARs tilt angles in every $5^{\circ}$ interval within the area ranges $<$50 millionth, 50--250 millionth , 250--500 millionth, and $>$500 millionth, respectively. The error bars represent the 95$\%$ confidence interval, whilst the circles and square symbolize the signs of average title angles. The color red and blue are used to indicate the tilt angle in the northern and southern hemispheres, respectively. The blue vertical lines represent the cumulative magnetic flux in every $5^{\circ}$ interval.}
    \label{fig13}
\end{figure*}

\begin{figure*}[ht!]
    \centering
    \includegraphics[width=0.8\textwidth]{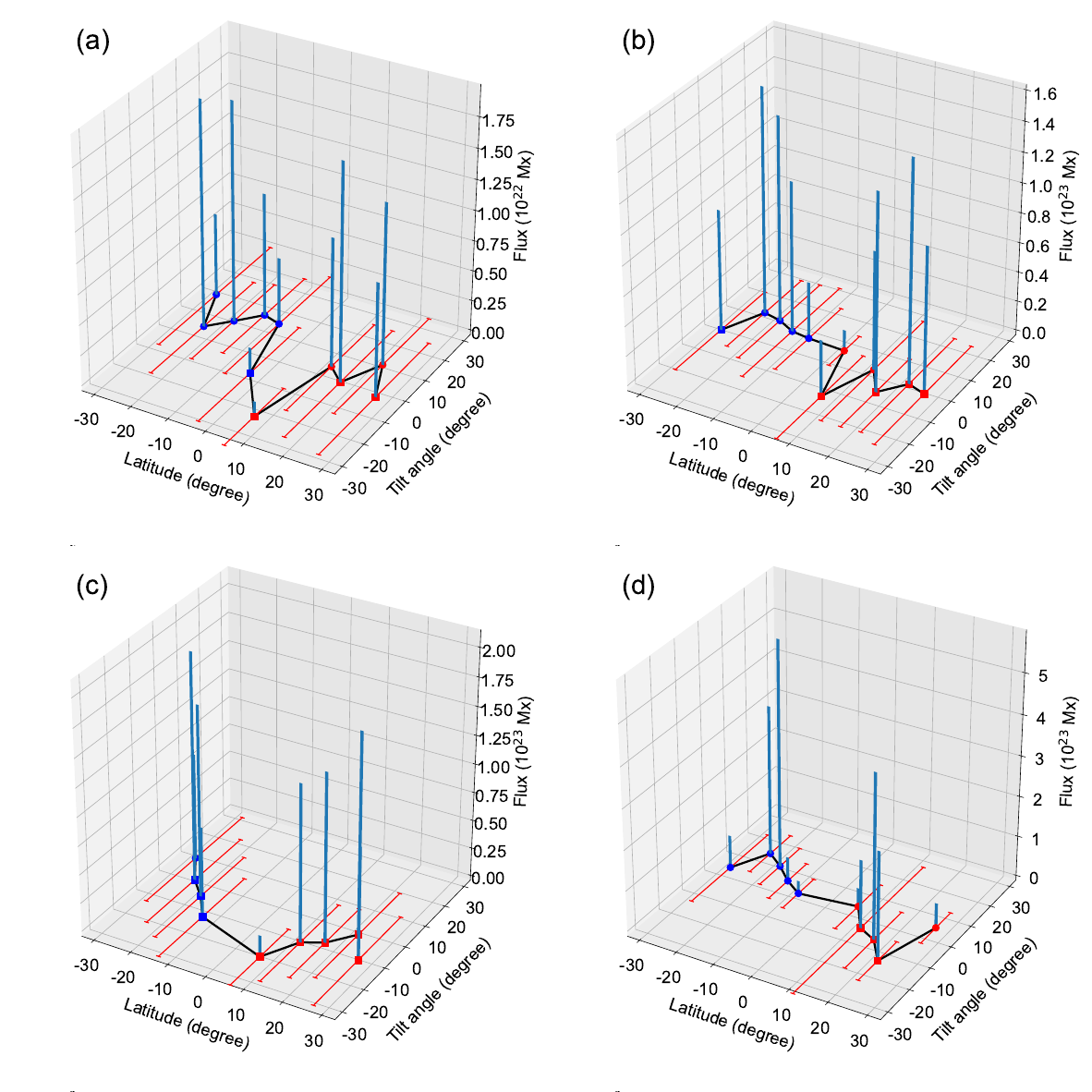}
    \caption{Similar to Figure~\ref{fig13} but for the rising phase of solar cycle 25.}
    \label{fig14}
\end{figure*}

The systematic tilt of sunspot groups with respect to the east–west direction was first identified by \citet{Hale1919}. The increase of the average tilt angle with latitude—where the leading spot is closer to the equator than the following spot—was termed ``Joy's Law.'' The study of active region tilt angles and their long-term variation plays a crucial role in understanding the subsurface dynamics of magnetic flux and the dynamo mechanism \citep{Bhowmik2018}. With the high-resolution magnetograms from SOHO/MDI and SDO/HMI, the characteristics of the magnetic tilt angles of bipolar regions have been extensively analyzed \citep{Driel2015}. In our analysis, the magnetic tilt angle of an AR is defined as the angle between the line connecting the centers of the positive and negative polarities and the equator. The sign of the tilt angle is defined as positive if the latitude of the positive polarity region is larger than that of the negative polarity region, and negative if the opposite is true, applicable in both hemispheres.

In accordance with the aforementioned definition, the magnetic tilt angle of each bipolar AR is calculated. The tilt angles are divided into 18 groups at 10-degree intervals, and their yearly histograms are plotted in Figure~\ref{fig11}(a). The upward histogram represents the number of ARs with positive tilt angles, while the downward histogram corresponds to the number of ARs with negative tilt angles. It is seen that during solar cycle 24, the number of ARs with positive tilt angles exhibited two peaks: the first in 2011 and the second in 2013. In contrast, the number of ARs with negative tilt angles increased from 2011, peaked in 2014, and then decreased during the declining phase. This number increased again from 2020 to 2023 during the rising phase of solar cycle 25. 

Similarly, the histograms of the bipolar AR tilt angles at different latitude bands are shown for solar cycle 24 in Figure~\ref{fig11}(b) and the rising phase of solar cycle 25 in Figure~\ref{fig11}(c), respectively. Joy's Law is evident from the two panels as the histograms at larger latitudes have a longer tail, i.e., the tilt angle becomes larger as the latitude increases. During solar cycle 24, there are more ARs with positive tilt angles than those with negative tilt angles in the northern hemisphere, while the opposite is true in the southern hemisphere. This pattern reversed during the rising phase of solar cycle 25.

To clearly illustrate the evolution of the tilt angle distribution, we plot the tilt angles at different latitude bands with vectors, which are overlaid on the butterfly diagram of ARs, as illustrated in Figure~\ref{fig12}. The red and black dots represent the positive and negative polarities of bipolar ARs, respectively. Each vector indicates the average tilt angle at specific latitude and year. It is noted that the averaged tilt angle is so small that the difference would be indistinguishable. To make the difference clearer, the orientation of the vectors is amplified by 8 times. From the plot, a discernible pattern consistent with Joy's law is evident for ARs at latitudes lower than $30^{\circ}$, where the leading polarities are closer to the equator, while the following polarities are farther from the equator. It is also evident that the tilt angles of ARs at relatively high latitudes exhibit a marked increase, resulting in a significant disparity compared to those at lower latitudes. Notably, an anti-Joy's law pattern is observed in ARs in the southern hemisphere during the declining phase of solar cycle 24, spanning from 2017 to 2019.

As indicated by Joy's law, the tilt angle increases with latitude. This phenomenon has been confirmed by numerous authors through various observations \citep{Driel2015}. We calculate the latitudinal variation of the average tilt angle in different latitude bands at $5^{\circ}$ intervals in both hemispheres during solar cycle 24 and the rising phase of solar cycle 25, respectively. The results are shown in Figures~\ref{fig13} and \ref{fig14}. Panels (a)--(d) in Figure~\ref{fig13} illustrate the latitudinal distribution of the average tilt angles of ARs in the area ranges $<$50 millionths, 50--250 millionths, 250--500 millionths, and $>$500 millionths during solar cycle 24, respectively. The blue vertical lines represent the cumulative magnetic flux in each $5^{\circ}$ interval. ARs with areas smaller than 50 millionths demonstrate random tilt variations. By considering the cumulative magnetic flux across different latitudes, it is found that the average tilt angle in the latitude bands $5^{\circ}$ to $15^{\circ}$ in the northern hemisphere, with relatively large cumulative magnetic flux, follows Joy's law. A similar variation is observed for ARs in the area range of 50--250 millionths, but in the southern hemisphere. However, the average tilt angle in the latitude bands $5^{\circ}$ to $15^{\circ}$ in the northern hemisphere, with relatively large cumulative magnetic flux, deviates from Joy's law. An opposite tilt angle pattern is observed for ARs in the area range of 250--500 millionths. The average tilt angle of ARs with areas exceeding 500 millionths in the latitude bands $5^{\circ}$ to $15^{\circ}$ adheres to Joy's law. ARs at relatively low or high latitudes exhibit random average tilt angles with relatively low cumulative magnetic flux across all area ranges. 

As shown in panels (a) and (b) of Figure~\ref{fig14}, during the rising phase of solar cycle 25, the average tilt angles of ARs with areas smaller than 250 millionths show random variations, except for those in the area range of 50--250 millionths, which follow Joy's law from $5^{\circ}$ to $25^{\circ}$. The average tilt angle of ARs in the area range of 250--500 millionths decreases with increasing latitude in the latitude band $5^{\circ}$ -- $20^{\circ}$, violating Joy's law. However, larger ARs with areas exceeding 500 millionths follow Joy's law in the latitude band $5^{\circ}$ -- $20^{\circ}$. 

The error bars indicating 95\% confidence for the average tilt angle at different latitudes are plotted in Figures~\ref{fig13} and \ref{fig14}. In both solar cycle 24 and the rising phase of solar cycle 25, the distribution of error bars across different area ranges exhibits a similar pattern: for ARs with areas smaller than 500 millionths, the error bars are very large, often exceeding the range of the average tilt angle distribution at all considered latitudes. In contrast, for ARs with areas larger than 500 millionths, the error bars are relatively smaller, providing greater confidence that Joy's law is satisfied in the latitude range of $5^{\circ}$ -- $20^{\circ}$.

\begin{figure*}[ht!]
    \centering
     \includegraphics[width=\textwidth]{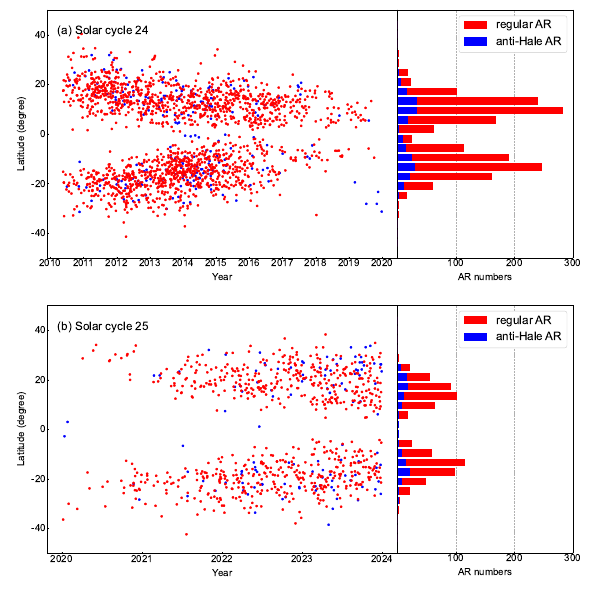}
    \caption{Latitudinal distribution of ARs categorized by whether they violate Hale's law. (a) Latitudinal distribution of regular and anti-Hale ARs during solar cycle 24, with red and blue dots representing regular and anti-Hale ARs, respectively. The histogram on the right side of the figure presents the distribution of AR numbers for regular and anti-Hale ARs in different latitude bands. Panel (c) is similar to (b) but for those during the rising phase of solar cycle 25.}
    \label{fig15}
\end{figure*}

\subsection{Distributions of the Regular and Anti-Hale Active Regions}

Hale's law states that the leading polarity of bipolar ARs is negative in the northern hemisphere and positive in the southern hemisphere during even solar cycles, and vice versa during odd solar cycles \citep{Hale1919}. The detected ARs in this study conform to Hale's law, as shown in Figure~\ref{fig12}. Specifically, the leading polarities are negative in the northern hemisphere and positive in the southern hemisphere during solar cycle 24, while the opposite is observed in the rising phase of solar cycle 25. However, it is important to note that some ARs exhibit distinct morphological characteristics and violate Hale's law. These so-called ``anti-Hale'' ARs have been reported by numerous authors \citep{Stenflo2012, Driel2015, Li2018, Zhukova2020, Jaramillo2021, Zhukova2024}. In this study, we identify both regular and anti-Hale ARs among the detected ARs and plot the results in Figure~\ref{fig15}. The red and blue dots represent the latitudes of regular and anti-Hale ARs, respectively. As shown in Figure~\ref{fig15}(a), anti-Hale ARs are distributed across all latitudes below $40^{\circ}$ and throughout the entirety of solar cycle 24. A notable increase in the number of anti-Hale ARs was observed in 2014, particularly in the southern hemisphere, around the time of solar maximum. The latitudinal distribution of both regular and anti-Hale ARs exhibits a bimodal pattern, with peaks between $10^{\circ}$ and $20^{\circ}$ in both hemispheres. A similar trend is observed during the rising phase of solar cycle 25. It is revealed that the proportion of ARs violating Hale's law is 13\% and 16\% in solar cycles 24 and 25, respectively.

\section{Discussion and Conclusions}\label{discussion&conclusion}

In this study, an automated detection method for solar ARs in LOS full-disk magnetograms was developed. The method, based on the DBSCAN clustering algorithm, was named DBSCAN$^2$-based Solar Active Region Detection (DSARD). It is an unsupervised machine learning method that does not require pre-labeled active regions as training samples. The validity of the method was established by using the NOAA catalogue as the reference standard. The average true positive rate ($R_{TP}$) and false positive rate ($R_{FP}$) of the method were determined to be $91.8\%$ and $7.2\%$, respectively. Subsequently, LOS full-disk magnetograms observed by SDO/HMI from May 2010 to December 2023 were processed using this method to investigate the long-term variation of ARs. To avoid mulitple detections and to minimize the projection effects, we examined the ARs only within a longitudinal range of $\pm 6^{\circ}$ from the central meridian of the solar disk. A total of 1991 ARs were detected within this range during solar cycle 24, and 872 ARs were detected during the rising phase of solar cycle 25. The evolution of these active regions presented the typical butterfly diagram. We calculated the drift velocities, the location, area, magnetic flux, and tilt angles of these ARs across different years and latitude bands.

A strong correlation exists between the yearly variation of AR numbers detected by our method and those recorded in the NOAA catalogue from 2010 to 2023, although our method yields comparatively more ARs. This discrepancy can be attributed to the enhanced sensitivity of our method, which enables detecting relatively small and diffuse ARs. Our analysis revealed the presence of two peaks: the first occurs in the northern hemisphere around 2012, and the second appears in the southern hemisphere around 2014. This finding aligns with observations of sunspots reported in previous studies \citep{Chowdhury2019, Veronig2021, Nandy2021, Du2022, Andreeva2023}. During the rising phase of solar cycle 25, the number of ARs increased more rapidly than during the early phase of solar cycle 24, suggesting that the strength of solar cycle 25 may exceed that of solar cycle 24. The latitudinal distribution of AR numbers exhibits a bimodal pattern, with a majority of ARs concentrated within the latitude band of $10^{\circ}$--$20^{\circ}$ and few above $40^{\circ}$. 

The latitudinal migration of ARs is illustrated in the butterfly diagrams shown in Figure~\ref{fig6} for solar cycle 24 and the rising phase of solar cycle 25. Relatively large ARs began to appear in the northern hemisphere around 2012, while in the southern hemisphere, they emerged in late 2012 and became particularly large and numerous in 2014 and 2015. This indicates that the ARs in the southern hemisphere were stronger than those in the northern hemisphere during the solar maximum of solar cycle 24. In solar cycle 25, the solar activity ascended more rapidly, and opposite to the previous cycle, ARs are dominant in the southern hemisphere.
 
To quantitatively analyze the migration of ARs during solar cycle 24 and the rising phase of solar cycle 25, the monthly average latitudes of ARs were calculated, and a cubic polynomial fitting was applied to derive their drift velocities. During solar cycle 24, ARs in the southern hemisphere exhibited relatively stable drift velocities, ranging between [0.7, 0.8] m s$^{-1}$. In contrast, the drift velocities of ARs in the northern hemisphere showed more pronounced variations. They initially decreased from 1.5 m s$^{-1}$ to 0.1 m s$^{-1}$ in late 2015, before increasing to 0.8 m s$^{-1}$ by the end of solar cycle 24 in 2019. During the rising phase of solar cycle 25, clear asymmetries in drift velocities were also observed between the two hemispheres. In the northern hemisphere, the drift velocities decelerated towards higher latitudes from 2021 to 2022, followed by an acceleration towards the equator, reaching 1.0 m s$^{-1}$ in 2023. Conversely, the southern hemisphere exhibited a uniform acceleration of drift velocities towards the equator, increasing from 0.5 m s$^{-1}$ to 1.2 m s$^{-1}$.
 
\citet{Li2010} analyzed the drift velocities of sunspot groups from 1919 to 1989, covering six solar cycles recorded by the Royal Greenwich Observatory and NOAA. They identified a two-stage migration pattern: sunspot groups initially migrate from latitudes of approximately $28^{\circ}$ with a drift velocity of about 1.2 m s$^{-1}$ towards the solar equator from the beginning of a cycle to the cycle maximum. This is followed by a transition to a drift velocity of about 1.0 m s$^{-1}$ at latitudes of approximately $20^{\circ}$ towards the solar equator during the remainder of the solar cycle, ultimately reaching a final latitude of about $8^{\circ}$. The results obtained in this study demonstrate that the migration of ARs also exhibited a two-phase pattern within solar cycle 24: from the beginning to the solar maximum, the drift velocity decelerates, followed by an acceleration during the declining phase. \citet{Zhang2010} applied their automated detection method to identify ARs in synoptic magnetograms constructed by SOHO/MDI from 1996 to 2008 during solar cycle 23. They found a linear drifting mode and derived an average drift velocity of 0.708 $\pm$ 0.015 m s$^{-1}$. The drift velocities derived from the ARs detected in the present study are consistent with those reported in previous studies, although minor discrepancies may arise due to differences in data sources and time ranges.

The AR areas were divided into four groups based on their sizes: $<$50 millionths, 50--250 millionths, 250--500 millionths, and $>$500 millionths, with the proportions being approximately $1:2:4:2$. The latitudinal distribution of the AR numbers in the four area groups, as well as their cumulative areas during solar cycle 24 and the rising phase of solar cycle 25, exhibited bimodal patterns. Only a few ARs are located in the latitude band above $30^{\circ}$ in both hemispheres. The peaks of AR numbers occurred in the latitude band of $10^{\circ}$--$20^{\circ}$, with the exception of ARs in the area range of 50--250 millionths in the northern hemisphere and those in the range of 250--500 millionths in the southern hemisphere. Notably, the AR numbers in the latitude band of $30^{\circ}$--$40^{\circ}$ during the rising phase of solar cycle 25 were comparatively larger than those in solar cycle 24. This suggests that ARs were distributed over a wider range of latitudes in solar cycle 25.

The latitudinal distribution of the AR numbers in the four area groups and their cumulative area in solar cycle 24 and the rising phase of solar cycle 25 manifested as bimodal distributions, with a few ARs occupying the latitude band over $30^{\circ}$ in both hemispheres. The peaks of AR numbers appeared in the latitude band $10^{\circ}$--$20^{\circ}$, with the exception of ARs with area within 50--250 millionth in the northern hemisphere and that with area within 250--500 millionth in the southern hemisphere. It is noteworthy that the AR numbers in the latitude band $30^{\circ}$--$40^{\circ}$ during the rising phase of solar cycle 25 were comparatively larger than those in solar cycle 24. This observation suggests that the ARs were distributed over a wider range of latitudes in solar cycle 25.

The magnetic flux was calculated by summing the magnetic fluxes in each AR, yielding the positive, negative, and total unsigned fluxes for each individual AR, respectively. Similar to AR areas, the AR magnetic fluxes were also divided into four groups based on their intensities: $< 10^{21}$ Mx, 1--5 $\times10^{21}$ Mx, 5--10 $\times10^{21}$ Mx, and $>10^{22}$ Mx. It was observed that the magnetic flux exhibits a yearly and latitudinal distribution consistent with that of the AR areas, indicating a correlation between the two variables. The net magnetic flux was found to be negative from 2010 to 2013, becoming positive with relatively large values in 2014, concurrent with the polar reversal in solar cycle 24. Subsequently, the net magnetic flux remained near zero during the decay phase of solar cycle 24. During the rising phase of solar cycle 25, the positive and negative magnetic fluxes exhibited rapid growth. However, the net magnetic flux remained approximately zero. The latitudinal distribution of the cumulative net magnetic flux of ARs revealed that, in solar cycle 24, the net magnetic flux is positive in the southern hemisphere and negative in the northern hemisphere within the relatively low latitude bands of $0^{\circ}$ to $30^{\circ}$. Conversely, during the rising phase of solar cycle 25, the net magnetic flux exhibits a reverse trend, albeit with comparatively weaker strength.

The asymmetry indices of the AR number, cumulative area, and total unsigned magnetic flux revealed that the northern hemisphere was dominant across all types of ARs, with the exception of the periods 2013–2014 and 2020–2021. Notable asymmetry indices were also observed in 2012, 2015, and 2022. The asymmetry indices for cumulative positive and negative magnetic fluxes exhibited similar trends, albeit with some exceptions. These indices suggest that the northern hemisphere dominated in ARs with both positive and negative magnetic flux during the rising and decay phases of solar cycle 24, while the southern hemisphere becomes dominant around the solar maximum. During the rising phase of solar cycle 25, northern hemisphere was dominant for the cumulative positive magnetic flux, whereas the opposite is true for cumulative negative magnetic flux. However, these asymmetry indices are not pronounced.

The magnetic tilt angle of each bipolar AR was calculated, and its distribution was analyzed with respect to years and latitude bands. The proportion of ARs with large tilt angles increased with latitude, a phenomenon consistent with Joy's law. To provide a clear and intuitive representation of the tilt angle distribution, the mean tilt angles of bipolar ARs in different latitude bands were plotted and overlaid on the butterfly diagram of ARs. The long-term variation of tilt angles is presented in Figure~\ref{fig12}. A clear pattern consistent with Joy's law is evident for the ARs at latitudes lower than $30^{\circ}$. At higher latitudes, the tilt angles of ARs increase, accompanied by a widening disparity compared to those at lower latitudes. Notably, an anti-Joy's law pattern was observed in the southern hemisphere as solar cycle 24 approached minimum, spanning from 2017 to 2019.

\citet{Tlatov2013} discovered that bipolar ARs with areas exceeding 300 millionths of the solar hemisphere (MHS) generally follow Joy's law, whereas those with areas in the range of 50 to 300 MHS exhibit an inverse pattern to Joy's law. In this work, a similar tendency to follow Joy's law was identified in relatively large ARs with areas over 500 millionths, particularly within the high cumulative magnetic flux latitude band of $5^{\circ}$ to $20^{\circ}$. In contrast, smaller bipolar ARs displayed random variations in their average tilt angles. The observed randomness in the results can be attributed to the wide error bars associated with small bipolar ARs. These large error bars are likely due to significant fluctuations at all scales \citep{Stenflo2012}. Furthermore, \citet{Tlatov2013} reported a bending of Joy's law at $30^{\circ}$ for bipolar ARs with areas exceeding 300 MHS, while those with areas between 50 and 300 MHS exhibited bending at latitudes ranging from $30^{\circ}$ to $40^{\circ}$. Regarding the findings of this study, it is noteworthy that, if the influence of error bars is disregarded and the focus is placed solely on the average tilt angle, deviations from Joy's law are observed. Specifically, large ARs (with areas over 500 millionths) show deviations at $20^{\circ}$ latitude, while relatively small ARs (with areas between 50 and 250 millionths) in the southern hemisphere exhibit deviations at $25^{\circ}$ latitude.

The detected ARs exhibited strong adherence to Hale's law, with leading polarities being negative in the northern hemisphere and positive in the southern hemisphere during solar cycle 24. This pattern reversed during the rising phase of solar cycle 25. However, it is important to note that a subset of ARs violated Hale's law, accounting for 13\% and 16\% of all ARs during solar cycle 24 and the rising phase of solar cycle 25, respectively. \cite{Zhukova2020} reported that approximately 3.0\% of all studied ARs were anti-Hale regions, based on their selection criteria. In contrast, recent studies found higher percentages of anti-Hale ARs, ranging from 4\% to 8\% \citep{Khlystova2009,McClintock2014,Li2018}. Notably, the findings of our study, reaching 13\% and 16\%, align with these recent results. \cite{Zhukova2020} proposed several reasons that may explain these observations. First, small, short-lived ARs are more likely to exhibit anti-Hale patterns. Second, the enhanced sensitivity of modern instruments enables the detection of weaker ARs. Third, identifying anti-Hale ARs within activity complexes remains a significant challenge. Additionally, the high sensitivity of the DSARD method in detecting ARs may also contribute to the observed results. The distribution of anti-Hale ARs spans all latitudes below $40^{\circ}$ and extends throughout the entirety of solar cycle 24. A notable increase in the number of anti-Hale ARs was observed in 2014, particularly in the southern hemisphere, coinciding with the solar maximum. The latitudinal distributions of both regular and anti-Hale ARs exhibit a bimodal pattern, with peaks occurring in the latitude bands of $10^{\circ}$ to $20^{\circ}$ in both hemispheres. It is worth noting that similar trends were observed during the rising phase of solar cycle 25. 

In summary, an automated detection method for solar ARs in LOS full-disk magnetograms was developed. This method is based on the DBSCAN clustering algorithm. The validity and accuracy of the method was demonstrated through comparison with the NOAA catalogue. The method was then applied to process the LOS full-disk magnetograms observed by SDO/HMI from May 2010 to December 2023 to obtain the long-term variation of ARs. The key results are summarized as follows:

\begin{enumerate}
	\item The total number of ARs peaked around 2014, with the maximum number occurring in 2012 in the northern hemisphere and in 2014 in the southern hemisphere. The total area and total magnetic flux reached their maxima in late 2014. Approximately 99\% of ARs were located at latitudes below $40^{\circ}$, with their latitudinal distribution peaking in the $[10^{\circ}, 20^{\circ}]$ band in both hemispheres.
	
	\item The latitudinal distributions of the total number, area, and magnetic flux of ARs were predominantly bimodal, with peaks in the $[10^{\circ}, 20^{\circ}]$ latitude band in both hemispheres.
	
	\item The latitudinal migration of ARs during solar cycle 24 exhibited two distinct phases: from the beginning of the cycle to the year following the solar maximum, the drift velocity decelerated, then accelerated during the declining phase. The drift velocities ranged between $[0.1, 1.6] \, \text{m s}^{-1}$. Variations in drift velocities were significantly more pronounced in the northern hemisphere compared to the southern hemisphere, both during solar cycle 24 and the rising phase of solar cycle 25.
	
	\item The asymmetry indices of the AR number, cumulative area, and total unsigned magnetic flux all indicate that the northern hemisphere dominated in terms of AR activity, except during the periods 2013--2014 and 2020--2021.
	
	\item Bipolar ARs with latitudes below $30^{\circ}$ follow Joy's law. However, the tilt angles of ARs at latitudes above $30^{\circ}$ exhibit no clear pattern.
	
	\item ARs with areas smaller than 500 millionths showed random variations in the average tilt angles. In contrast, ARs with areas larger than 500 millionths and high cumulative magnetic flux in the $5^{\circ}$ to $20^{\circ}$ latitude band followed Joy's law, with a bending of Joy's law observed above the latitude $20^{\circ}$.
	
	\item Approximately 13\% and 16\% of all ARs violated Hale's law during solar cycle 24 and the rising phase of solar cycle 25, respectively.
\end{enumerate}




\begin{acknowledgements}
We are grateful to the SDO teams for providing the observational data. The present work was supported by the National Key Research and Development Program of China (2020YFC2201200), NSFC under grants 12173019 and 12127901, and the Young Data Scientist Program of the China National Astronomical Data Center, as well as the AI \& AI for Science Project of Nanjing University.
\end{acknowledgements}

\bibliography{manuscript_ar}{}
\bibliographystyle{aasjournal}



\end{document}